\documentclass[12pt]{article}

\usepackage{cite}

\def\jetset{{\sc Jetset}}

\def\C2q{C_2(Q)}
\def\tC2q{$\C2q$}
\def\f2q{f_2(Q)}
\def\tf2q{$\f2q$}

\def\lep2{LEP~2}
\def\mw4j{m_{\W}^{4j}}
\def\tmw4j{$\mw4j$}
\def\delq{\delta Q}
\def\tdelq{$\delq$}
\def\BE{\mbox{\sc BE}}

\newcommand{\PLB}[3]{Phys.~Lett.\ {\bf B#1} ({#3}) {#2}}

\newcommand{\bfp}{\mathbf{p}}
\newcommand{\bfr}{\mathbf{r}}

\newcommand{\e}{\mathrm{e}}
\newcommand{\p}{\mathrm{p}}
\newcommand{\q}{\mathrm{q}}
\newcommand{\W}{\mathrm{W}}
\newcommand{\Z}{\mathrm{Z}}
\newcommand{\pbar}{\overline\mathrm{p}}
\newcommand{\qbar}{\overline\mathrm{q}}
\def\ee2ww{$\e^+\e^-\rightarrow\W^+\W^-$}

\begin{document}

\sloppy

\begin{titlepage}

  \begin{flushright}
    LU TP 97--30\\
    NORDITA-97/75 P\\
    hep-ph/9711460\\
    November 1997
  \end{flushright}
  \begin{center}

    \vskip 10mm
    {\Large\bf Modelling Bose--Einstein correlations at LEP~2}
    \vskip 15mm

    {\large Leif L\"onnblad}\\
    NORDITA\\
    Blegdamsvej 17\\
    DK-2100 K\/obenhavn \/O, Denmark\\
    leif@nordita.dk
    
    \vskip 10mm

    {\large Torbj\"orn Sj\"ostrand}\\
    Dept.~of Theoretical Physics\\
    S\"olvegatan 14A\\
    S-223 62  Lund, Sweden\\
    torbjorn@thep.lu.se

  \end{center}
  \vskip 30mm
  \begin{abstract}

    We present new algorithms for simulating Bose--Einstein
    correlations among final-state bosons in an event generator. The
    algorithms are all based on introducing Bose--Einstein correlations 
    as a shift of final-state momenta among identical bosons, and differ 
    only in the way energy and momentum conservation is ensured. The
    benefits and shortcomings of this approach, that may be viewed as 
    a local reweighting strategy, is compared to the ones
    of recently proposed algorithms involving global event reweighting.
    
    We use the new algorithms to improve on our previous study of the
    effects of Bose--Einstein correlations on the W mass measurement at
    \lep2. The intrinsic uncertainty could be as high as 100 MeV but
    is probably reduced to the order of 30 MeV with realistic
    experimental reconstruction procedures.

  \end{abstract}

\end{titlepage}

\section{Introduction}

Most of the particles produced in hadronic events are pions, and as
such they obey Bose statistics. One therefore expects an enhancement
of the production of identical particles at small momentum separation,
relative to what uncorrelated production would have lead to
\cite{orig}.  The shape of the enhancement curve reflects the size of
the space--time region over which particle production occurs and the
mechanism of particle production. Measurements of Bose--Einstein (BE)
effects therefore directly test our understanding of QCD, in a way
very much complementary to other QCD studies.

Unfortunately, the nice basic idea has complications. We do not
have a solution of nonperturbative QCD even for the case of 
nonidentical particles, let alone for identical ones. Thus we do not
know how to write down the amplitudes that, when symmetrized, 
should lead to a BE enhancement. That is, theoretical studies have 
to be based on models, and so shortcomings in comparisons with
data may be difficult to localize. From the experimental point of
view, the extraction of an unbiased BE enhancement curve is 
impossible, since there is no access to an alternative world not
obeying BE statistics but otherwise the same. Reference samples
can be defined in various ways, but all suffer from limitations.
   
That notwithstanding, studies of multihadronic events show clear 
evidence of BE enhancements \cite{reviews,Hay,LEPdata}.
If the enhancement of the two-particle correlation is parametrized 
in the phenomenological form
\begin{equation}
f_2(Q) = 1 + \lambda \exp(-Q^2 R^2) ~,
\label{ftwo}
\end{equation}
one finds $\lambda \sim 1$ and $R \sim 0.5$~fm in hadronic 
$\e^+\e^-$ annihilation events. Here $Q$ is the relative difference 
in four-momenta, $Q^2 = Q_{12}^2 = - (p_1 - p_2)^2 = m_{12}^2 -4 m^2$.
The $\lambda \sim 1$ value refers to production at the primary
vertex; decays of long-lived resonances and other dilution effects
lead to the observable values typically being more like 0.2--0.3.
The $R$ parameter does not have to have a simple interpretation,
but can be identified with a source radius in geometrical models 
\cite{geom}.

One interesting question is whether BE correlations only affect our
understanding of QCD, or whether it has wider implications.
In a previous publication \cite{BE95} we investigated possible
BE effects on the W-mass measurement at \lep2.
Such effects can be expected in the purely hadronic channel because
the space--time regions of hadronization of the two W bosons are
overlapping. Using an algorithm which models BE correlations in the
{\sc Pythia} \cite{pythia} event generator in terms of a `final-state
interaction' between identical bosons, we found that the effects on the
measured mass in the purely hadronic channel, also called the four-jet
channel, \tmw4j, may be very large. Although the algorithm had some
shortcomings, it was the first serious attempt to estimate this effect
and still represents a thought-provoking `worst case' scenario
indicating a systematic uncertainty of more than 100 MeV on \tmw4j.

Since our first publication, several other studies have been performed 
\cite{Jadach,Rasmus,fial,Sharka,JariMarkus}, giving
small or vanishing effects on \tmw4j. Contrary to our approach, these
new algorithms are mainly based on a global reweighting of events to 
obtain the observed correlations between identical bosons. It is often 
argued that such algorithms are more `theoretically appealing' than
the local reweighting perspective that is implicit in our momentum
shifting strategy. As we point out in \cite{BE95} and also stress in 
this paper, this need not be the case: the global reweighting philosophy 
can give unexpected and unphysical side effect. We cannot therefore 
today claim that there is one `best' recipe. As long as these 
uncertainties persist, we cannot exclude a significant systematic 
shift on \tmw4j.

It may, however, be possible to use other experimental observables
than \tmw4j to rule out one or several models. One such observable is
presented by DELPHI \cite{NoWWBE}. By a clever combination of
semi-leptonic and fully hadronic events, they can isolate the BE
effects due to correlations between pions from different W bosons. The
statistics is rather small, and so does not really discriminate between
models, but it is still interesting that DELPHI finds no trace of 
such BE effects. Recently ALEPH came to the same conclusion 
\cite{ALEPHno}. Should these results survive an increase in
statistics, it would require a revision of our current understanding of
such BE effects and would surely rule out a significant shift of \tmw4j
by this source. It would favour a scenario where the W$^+$ and W$^-$
systems appear as uncorrelated sources of particle production, in spite 
of their space--time overlap. While the (lack of) BE enhancement 
does not directly probe other possible sources of mass shifts, such as 
colour rearrangement \cite{TSVK,rearrange}, a null result would make it 
plausible that also these other sources are negligible. From J/$\psi$
production in B meson decay we know that the colour rearrangement 
mechanism does exist, however, so conclusions have to be drawn with 
care. 

The main problem with the the algorithm we presented in \cite{BE95} is
that energy conservation is explicitly broken in the treatment of
individual particle pairs, and is restored only by a global rescaling
of all final-state hadron momenta. This rescaling introduces an
artificial negative shift in \tmw4j, and a rather cumbersome
correction scheme is needed to unfold the positive shift due to BE
effects. Therefore it was not feasible to study the consequences of
realistic experimental reconstruction procedures. In this paper we
present four new algorithms, all variations of the same basic
`final-state-interaction' approach, where not only momentum but also
energy conservation is handled locally.  The algorithms are presented
in detail in section \ref{sec:alg}.  Before that, however, we have a
discussion in section \ref{sec:weight} on the understanding and
modelling of the BE phenomenon in general, to clarify some of the
conceptual issues, in particular the reasons for us to pick a local
approach to the BE phenomenon. In section \ref{sec:res} we present
some results using our new algorithms, and finally, we present our
conclusions in section \ref{sec:sum}.

\section{Models and data for the BE phenomenon}
\label{sec:weight}

As already emphasized in the introduction, we do not know how to
include the BE phenomenon in descriptions of hadron production in
high-energy interactions. In this sense, whatever is currently done 
has the character of `cookbook' recipes, and should be taken with a
pinch of salt. This does not mean that all approaches have to be 
put on an equal footing: the level of sophistication and the measure 
of internal consistency can easily vary between models. 

\subsection{Global vs.\ local BE weights}

A possible characterization of models is in terms of `global' and
`local'. In global models a BE weight $W_{\mathrm{BE}}$ can be
associated with each individual event. More precisely, it is assumed 
that a model exists for particle production in the absence of
Bose statistics, that can be used to draw an unbiased sample of
events. In order to include BE effects, each such unbiased event 
obtains a weight that is the ratio of the squared matrix elements
of the production process with and without BE, respectively. 
The art is then to derive as plausible matrix elements as possible,
so that the ratio can be evaluated with some confidence. The hope is
that a lot of our ignorance should divide out in the ratio, so that we 
do not need absolute knowledge of nonperturbative QCD to make some
realistic predictions for $W_{\mathrm{BE}}$.

The word `global' is used to denote the character of the weighting 
procedure, in the sense that one weight is assigned to the event
as a whole, rather than to a specific particle pair.
The terminology is not intended to reflect the character
of the BE phenomenon as such, which normally is assumed to be local
in $(\Delta x, \Delta p)$ space. Thus the global weight is typically 
built up as the product or sum of factors/terms that each by itself 
is of local character. The introduction of a global weight still
leaves the door open for intentional or spurious BE effects of a
non-local character; e.g., the strength of the BE enhancement in one
region of an event could be influenced by the total multiplicity in
the rest of the event.

A global weight can be given different interpretations. Often it is 
viewed as a multiplicative factor affecting the production rate of a
given final state. In such approaches, there are some well-established 
experimental facts that have to be taken into consideration. Main 
among those is that the width of the Z$^0$ resonance agrees extremely 
well with the perturbative predictions of the standard model
\cite{precision}. If indeed there is a global BE weight 
$W_{\mathrm{BE}}$ for each event, such that
\begin{equation}
\Gamma_\mathrm{Z}^{\mathrm{total}} = \Gamma_Z^{\mathrm{leptonic}} +
\Gamma_\mathrm{Z}^{\mathrm{invisible} (\nu)} + 
\Gamma_\mathrm{Z}^{\mathrm{hadronic, perturbative}} \cdot 
\langle W_{\mathrm{BE}} \rangle
\end{equation}  
then $\langle W_{\mathrm{BE}} \rangle = 1$ to a precision much better
than 1\%. This immediately excludes models where weights always are
above unity, since a reweighting of events only at the per cent level
could not explain the order unity BE enhancements in the data.

Although precision is highest for $\Gamma_\mathrm{Z}$, some other
related conclusions can be drawn from other data. The 
$\langle W_{\mathrm{BE}} \rangle$ cannot be a function of energy, since 
$R = \sigma(\e^+\e^- \to \mathrm{hadrons})/\sigma(\e^+\e^- \to \mu^+\mu^-)$
agrees with perturbative predictions over a wide range of energies.
It also cannot be a function of initial quark flavour, since the b quark 
fraction of Z$^0$ decays agrees with electroweak theory. 
It appears implausible that BE weights could change the relative 
composition of partonic states, since both the distribution in number
of jets and in angles between jets agree very well with perturbative
QCD predictions, also when based on an $\alpha_{\mathrm{s}}$ determined
from other processes. In passing, we note that BE effects among the 
perturbative gluons are significantly reduced by the existence of
eight different colour states and are expected to be negligible.

Finally, the hadronic
multiplicity varies as a function of energy and primary flavour, so
the weight cannot be a function of the multiplicity in a direct way.
Implicitly it would still be, of course, in the sense that a larger
multiplicity for fixed energy and flavour means particles are packed
closer in phase space on the average, i.e.\ pairs have lower $Q$ values.
The increase of the average multiplicity with energy could then be
viewed as reflecting an increase in the phase space available for 
particle production, with unchanged average particle density in this 
phase space \cite{fractal}.

As we shall see, several models based on global weights have difficulties 
in accommodating these experimental observations.  From a theoretical
point of view, all the observations are naturally explained by them
having a common origin in the factorization property of QCD
\cite{factor}. Simply put, factorization tells that nonperturbative
physics cannot influence the hard perturbative phase, or at least that
any such corrections have to be suppressed by powers of $1/Q^2$, where
$Q$ here denotes the energy scale of the perturbative process. This
may be viewed as a natural consequence of the time-ordering of the
process, where first the Z$^0$ decays to a $\q\qbar$ pair, which then
may emit further partons that stretch confining colour fields, strings
\cite{AGIS}, between themselves.  The hadron production from the
string pieces only occurs at time scales of a few fermi in the center
of the event, and even later for the faster particles. By this time it
is `too late' to influence the original selection of q flavour or
(early) partonic cascade, but instead the hadronization process is
likely to proceed with unit probability to some final state.

Whereas many models with global weights break factorization, the ones 
with local weights take factorization as their starting point. A parton
configuration, once given by the perturbative rules, is fixed.
Any weighting that enhances some fragmentation histories must, 
in exact balance, deplete others with the same parton configuration.
Furthermore, the $R \sim 0.5$~fm value indicates that the BE effect
occurs predominantly on a local scale, affecting particles that
are produced fairly nearby along the string. Therefore, in the 
local models, it is assumed that the hadronization at one end
of the string occurs (almost) independently of that at the other 
end. This is already part of the standard string fragmentation 
approach, without BE, as a natural consequence of causality.
The acausality effects of the BE phenomenon are assumed to spread 
over distances of the order of $R$, in reality maybe some few fm,
but still small compared with the total size of the fragmenting
system at LEP energies. It is therefore assumed meaningless to define
a weight that attempts to bring together information about widely
separated parts of the event. Instead the local weight strategy is 
based on applying a reweighting procedure for each pair of identical
particles in a way that only affects the local neighbourhood of the 
pair. In practice, the BE phenomenon becomes reduced to a kind of
final-state interaction: the BE reweighting is a modest perturbation 
on events that, by and large, are given by the no-BE scenario.
This does not have to mean that underlying physics is 
that of a final-state interaction, only that the algorithms for local
weights can be made more tractable when reformulated in those terms.
Specifically, events generated without BE effects can be perturbed, 
by shifts in the momenta of the particles, in such a way as to give 
the desired two-particle correlations \cite{tsorig,BE95}. This procedure 
can be applied event by event, with unit probability.
 
It should be clear to the reader that we lean towards the local weigh
approach rather than the global weight one, since we do take the
experimental data and theoretical dogma of factorization seriously.
However, having said that, it must be admitted that the principles
of local weights does leave room for alternative and arbitrary
choices, e.g.\ as to how energy and momentum is conserved locally.
It is this arbitrariness that will be studied in the subsequent
sections. The global weight approach does not have the corresponding 
problem, since the reweighting is automatically between configurations
that all have the same energy and momentum. Currently the choice is 
therefore between the global models, that have a more appealing 
implementation but often contradict our current understanding of QCD, 
and the  local ones, that have a more sound basis in the factorization
properties of QCD but lead to rather ugly technical tricks. 
The distance between the ideal model and the algorithms actually
used may therefore be larger in the local approach. Specifically, what 
is studied in this paper is a set of local algorithms rather than the 
local concept as such.

It is possible to construct models intermediate to the pure `global'
and `local' extremes. In one existing model \cite{Sharka} factorization
is ensured by always retaining a parton configuration, once it has been
selected according to the perturbative rules. Only the subsequent 
hadronization step is assigned a weight, and repeated until accepted 
by standard Monte Carlo procedure. Also BE effects in decays are 
considered separately from the main reweighting loop. Thus the global 
weight aspects are minimized.

\subsection{Multiplicities}

A measure of our ignorance of the BE phenomenon is that we do not know 
whether it is supposed to change the multiplicity distribution of events 
or not. That is, does the `BE bump' at small momentum separation $Q$ 
values correspond to an extra number of particles in the event, that 
would not have been there in a world without Bose statistics? In thermal
field theory one can prove that $f_2(Q) \geq 1$ everywhere \cite{tft}, 
which would indicate that BE indeed does increase the average 
multiplicity, or at least changes the multiplicity distribution to
favour the high-multiplicity tail. However, the field theoretical 
definition of $f_2(Q)$ cannot be directly applied to $\e^+\e^-$
events, so already for this reason it is difficult to draw any
conclusions. Furthermore,
one of the necessary assumptions is that extra particles can
be produced at no cost in energy/momentum/charge/flavour conservation.
This may be a sensible approximation for the central rapidity region 
of heavy-ion collisions at very high energies (and even so it turns 
out to be problematical to implement BE models \cite{tftproblem}), 
but has little to do with our understanding of physics in $\e^+\e^-$ 
annihilation. Rather, a model like the string one implies that
particle production is based on local flavour conservation, so
that e.g.\ two positively charged particles could not appear as
nearest neighbours in rank. The string tension of 1 GeV/fm also sets 
the scale for how closely particles can be produced. There is 
therefore no logical need to assume a BE change of multiplicity. 
Just like ordinary fragmentation contains multiplicity fluctuations, 
however, one could imagine that the BE mechanism favours the 
fluctuations towards higher multiplicities; this is particularly
compelling in scenarios with global BE weights always above unity.

The data does not settle the issue. As conventionally presented, the
BE enhancement at small $Q$ is compensated by a dip of $C_2(Q)$ below
unity at intermediate $Q$. (In the following, we use $C_2(Q)$ for the
measured two-particle correlation and $f_2(Q)$ for the theory input.)
This behaviour is well `predicted' in our momentum shift algorithm,
i.e.\ it involves no free parameters but comes from the formalism. In
this sense, there is no case for a multiplicity change. However,
experimental analyses are normally based on a reference sample for the
imagined no-BE world picked to have the same multiplicity as the data.
By definition, one thus assumes no multiplicity change, and the dip at
intermediate $Q$ is a logical consequence of this assumption. In
model-independent fits, it is necessary to include a factor like $N(1
+ k Q)$ (with $k>0$ and $N < 1$), in addition to the form of
eq.~(\ref{ftwo}), to describe the data. Such a factor has no simple
interpretation in formalisms based on global weights always above
unity. However, if one plays with the main `$b$' parameter of the Lund
longitudinal fragmentation function \cite{AGIS} to create a Monte
Carlo no-BE reference world with a lower-than-real multiplicity, the
need for the $N(1 + k Q)$ factor vanishes for a multiplicity
$\sim$12\% lower than the data \cite{tompriv}. The $C_2(Q)$ still
drops below unity at very large $Q$, but this is an inevitable
consequence of energy conservation and not in contradiction with
weights always above unity. Finally, models with global weights both
above and below unity can explain the experimental dip at intermediate
$Q$ as part of the weight variation but, depending on the details of
the weight distribution, could additionally need to invoke some global
multiplicity change. Any answer between 0 and $\sim$12\% multiplicity
change thus seems perfectly feasible to accommodate from an
experimental point of view, depending on the model used to interpret
the data.

One should also note what is {\em not} found in the data. The BE 
effect, especially for BE weights assumed everywhere above unity, 
could be expected to lead to `runaway' situations where an
event or a region of an event consists almost entirely of $\pi^0$'s or 
$\pi^{\pm}$'s, since this would maximize the event weight. No signals 
for larger-than-expected fluctuations of this nature have been found
in the data, indicating that the no-BE picture of uncorrelated flavour 
production at adjacent string breakup vertices (modulo some technical 
complications included in realistic event generators) is a good first
approximation. However, we would welcome further studies, to quantify 
how big such effects could still be allowed by the data.

A perfectly plausible scenario is thus that BE effects
do not change the particle number or composition of events, but only
relative momentum separation between particles. This is the assumption 
pursued in our local scenarios.

\subsection{Local approaches}
\label{sec:local}

Above we have argued for a local scenario, wherein all the major
properties of the event can be given without any reference to the
BE phenomenon. The BE effect is then introduced as a perturbation.
This gives a large formal similarity with final-state interactions,
although the underlying physics may well be different. Anyway, this
similarity allows for a more tractable approach to the simulation
of BE effects. 

The algorithm presented in \cite{BE95} takes the hadrons produced by
the string fragmentation in \jetset, where no BE effects are present,
and shifts slightly the momenta of mesons so that the inclusive
distribution of the relative separation $Q$ of identical pairs is
enhanced by a factor \tf2q, e.g.\ of the form of eq.\ (\ref{ftwo}). 
Making the ansatz that the original distribution in $Q$ is just given 
by phase space, $d^3p/E\propto Q^2dQ^2/\sqrt{Q^2+4m^2}$, an appropriate 
shift $\delta Q$ for a given pair with separation $Q$ can be given by
\begin{equation}
  \int_0^Q\frac{q^2dq}{\sqrt{q^2+4m^2}} =
  \int_0^{Q+\delq}f_2(Q)\frac{q^2dq}{\sqrt{q^2+4m^2}}.
  \label{qshift}
\end{equation}
For an arbitrary $\f2q\geq 1$, \tdelq\ is negative and pairs are
pulled closer together. The pair density does not increase as fast
as phase space implies once $Q$ is larger than the typical transverse
momentum spread of the string fragmentation. This leads to the 
generated \tC2q dropping below unity at intermediate $Q$ and
approaching unity from below for large $Q$, see \cite{BE95} for 
details. The choice of not using the actual phase space density is a 
deliberate one; we believe that the deviations from a pure phase
space distribution of particles and the assumption of a conserved total
multiplicity should have repercussions in terms of the output \tC2q
not agreeing with the input \tf2q. 

The translation of \tdelq\ into a change in particle momenta is not
unique. Since the invariant mass of a pair is changed, it is not
possible to simultaneously conserve both energy and momentum, and so
compromises are necessary. We have chosen to conserve three-momentum
in the frame where the algorithm is applied. For a given pair of
particles $i$ and $j$ the change is $\bfp_i'=\bfp_i+\delta\bfp_i^j$,
$\bfp_j'=\bfp_j+\delta\bfp_j^i$, with
$\delta\bfp_i^j+\delta\bfp_j^i=\mathbf{0}$, and we simply take
$\delta\bfp_i^j=c(\bfp_j-\bfp_i)$ corresponding to pulling the
particles closer along the line connecting them in the current frame.
In \cite{BE95} we also tried other strategies, such as conserving
energy rather than momentum, and shifting the momenta of a pair in
their rest frame, but we found that our results were not very
sensitive to such choices.

A given particle is likely to belong to several pairs. If the momentum
shifts above are carried out in some specific order, the end result
will depend on this order. Instead all pairwise shifts are evaluated
on the basis of the original momentum configuration, and only
afterwards is each momentum $\bfp_i$ shifted to $\bfp_i' = \bfp_i +
\sum_{j \neq i} \delta \bfp_i^j$.  That is, the net shift is the
composant of all potential shifts due to the complete configuration of
identical particles. This means that the pair ansatz is strictly valid 
only for large source radii, when the BE-enhanced region in $Q$ is small, 
so that the momentum shift of each particle receives contributions only 
from very few nearby identical particles. For normal-sized radii, $R \sim
0.5$~fm, the method introduces complex effects among triplets and
higher multiplets of nearby identical particles, which (together with
the phase space ansatz discussed above) is reflected both in changes 
between the input \tf2q\ and the final output $C_2(Q)$ \cite{BE95,MSL} 
and in the emergence of non-trivial higher-order correlations. The latter 
actually agree qualitatively with such data \cite{DELPHIthree}.

Short-lived resonances like $\rho$ and K$^*$ are allowed to decay
before the BE procedure is applied, while more long-lived ones are 
not affected. This leads to a shift in the $\rho^0$ mass peak,
something also observed in the data \cite{rhodata}.

The above procedure preserves the total momentum, while the shift of
particle pairs towards each other reduces the total energy. For a
$\Z^0 \to \q\qbar$ event this shift is typically a few hundred MeV,
and so is small in relation to the $\Z^0$ mass. In practice, the
mismatch has been removed by a rescaling of all three-momenta by a
common factor (very close to unity). As a consequence, also the $Q$
values are changed by about the same small amount, whether the pairs
are at low or at high momenta. That is, the local changes due to the
energy conservation constraint have been minimized by spreading the
corrections globally.

By and large, the very simple ansatz above gives an amazingly good
account of BE phenomenology in $\e^+\e^-$ annihilation, including 
many genuine predictions. In addition to what has already been mentioned,
one could note the variation of longitudinal, out and sideways 
fitted radii as a function of the transverse mass of a pair 
\cite{mtdata}. Some of these agreements may be coincidental, or trivial 
consequences of any reasonable BE implementation, but at least 
$\e^+\e^-$ data so far has not revealed any basic flaw in the simple
original version of the local approach. 

By contrast, in $\p\pbar$ data the UA1 and E735 collaborations have 
observed that the $\lambda$ parameter decreases and the $R$ 
parameter increases with increasing particle density \cite{UAonedata}. 
Neither behaviour follows naturally from our approach, although it could 
be argued  that final-state interactions at least would be consistent
with an increasing radius of `decoupling' for larger multiplicities.
Above we have attempted to explain our momentum-shifting strategy
as being motivated more by a local reweighting philosophy than a 
final-state interaction one, in order to highlight similarities and
differences with global weight schemes. In view of the $\p\pbar$
data it might be prudent not to close the door on both effects being
present in the data, and hopefully both being approximated by our 
algorithm.

The agreement with $\e^+\e^-$ data does not mean that the method is 
free of objections \cite{Hay,fialcrit}. The
deterministic nature of the momentum shift algorithm does not go well
with the basic quantum mechanical nature of the problem, and is likely
to mean that a potential source of event-to-event fluctuations is lost. 
The selected input form of \tf2q, like in eq.~(\ref{ftwo}), is not 
coming from any first principles, and
$\lambda$ and $R$ are two free parameters. It could be argued that
$\lambda = 1$ is a natural value, and that a transverse BE radius 
$R \sim 0.5$~fm is about the transverse size of the string itself,
but it is not at all clear why a similar Gaussian form and radius should
apply for the longitudinal degree of freedom. This would require a 
detailed study and understanding of the microscopic history of the event 
(as is offered in some global models \cite{AndHof, AndRin, Sharka}).
Possibly it would then turn out that the shape used is reasonable
on the average, even when a poor approximation for the individual 
event. For instance, the space--time history of string fragmentation 
gives, on the average, a coordinate separation of two particle 
production vertices proportional to the momentum difference between 
the particles. The $Q^2$ factor of \tf2q could then be reinterpreted
as being $\Delta x \cdot \Delta p$, and the longitudinal $R$ related to 
longitudinal fragmentation parameters. However, the relation 
$\Delta x \propto \Delta p$ suffers from large fluctuations in the 
actual string histories, that are now completely neglected.

Another set of possible complications comes from the assumption that
the BE phenomenon is the same in quark and gluon jets, in spite
of the more complicated space-time structure of particle production
in the gluon jets, cf.\ the following model. Our local scheme is
here based on the simplest possible picture and, as for several of the
aspects covered above (spherical source, no input three-particle
correlation, \ldots), one could imagine more complicated variants 
of the local ansatz.   

\subsection{Global approaches}

Whereas the local approach to the BE phenomenon only has been
developed by us, many global algorithms have been proposed. It would
carry too far to describe all, but we here would like to 
comment on a few of them, with special emphasis on those that
have been used to study the issue of a \tmw4j shift. 

The probably most sophisticated global approach is the one originally
proposed by Andersson and Hofmann \cite{AndHof} and further developed 
by Andersson and Ringn\'er \cite{AndRin}. Here the fragmentation 
process is associated with a matrix element
\begin{equation}
\mathcal{M} = \exp (i \kappa - b/2) A ~,
\label{MBEbomarkus}
\end{equation}
where $\kappa$ is the string tension, $b$ is related to the breaking 
probability per unit area of the string and hence to the form of the
fragmentation function, and $A$ is the total space--time area spanned 
by the string before fragmenting. String histories with different areas
can lead to the same final state --- the simplest example being 
the permutation of the momenta of two identical particles --- 
so nontrivial interference effects are obtained when the amplitudes 
are added. This can be reformulated in terms of an effective weight
\begin{equation}
W_{\mathrm{BE}} = 1 + \sum_{\mathcal{P}' \neq \mathcal{P}}
\frac{ \cos \kappa \, \Delta A}%
{\cosh \left( \frac{\displaystyle b \, \Delta A}{\displaystyle2} + 
\frac{\displaystyle \Delta ( \raisebox{.5ex}{$\scriptscriptstyle \sum$} %
p_{\perp \q}^2)}{\displaystyle 2 \sigma^2} \right) } ~,
\label{WBEbomarkus}
\end{equation}
where $\mathcal{P}' \neq \mathcal{P}$ indicates that the sum should
run over all permutations of momenta of identical particles, except
for the original configuration itself. The second term in the
denominator comes from the transverse momentum degrees of freedom of
quark pairs that have to have their transverse momenta reinterpreted by
the permutation, and tends to dampen weights. The area difference 
$\Delta A$ between two string fragmentation histories is, for a simple 
pair permutation, equal to the product of the energy--momentum difference
and the four-distance between the production points. The cosine in
the weight numerator means that the $f_2(Q)$ distribution is expected
to oscillate around unity, while the dampening of the weight
denominator ensures that only the first peak and dip are visible in
the end.  The $\rho$ and $K^*$ decays are treated as if they were part
of the string decay itself, so that the decay products can be
symmetrized with primary particles. There are two technical
complications: firstly, that an inclusion of all possible permutations
would make the algorithm extremely slow and, secondly, that individual 
weights can be negative. The first point is ameliorated by a truncation,
where only terms with a significant impact on results are retained.
The latter point is an artifact of the algorithm and not a real problem.

The algorithm gives a good description of two-jet data, as far as it can 
be tested. However, it does give an average weight of about 1.2, that 
has to be divided out by hand. It is the oscillations of the weight 
function that gives it a value close to unity, with the actual number 
rather sensitive to fragmentation model parameters \cite{BoMarkuspriv}.
No clear physics interpretation is offered of the average weight,
e.g.\ in the context of the $\Z^0$ width. It has not been studied whether 
the algorithm gives a change in the jet number or primary flavour 
composition.

Technical complications means that the generalization of the model to 
three-jet events is less well studied. One consequence of the model 
is that a gluon jet is expected to contain less BE correlations than
a quark one: the gluon fragmentation involves two string pieces, so that
the distance between two particle production vertices, in absolute numbers
or defined in terms of $\Delta A$, is larger than implied by the 
momentum difference. In our local approach the full space--time 
hadronization history is not used, so this aspect is not caught.
Therefore one obtains differences between models, although they may be 
difficult to observe \cite{BoMarkuspriv}.

The model of Todorova--Nov\'a and Rame\v{s} \cite{Sharka} contains a 
global weight, but its importance is limited, so as to emphasize the
local character of the BE phenomenon. In a first step, a parton 
configuration is selected according to conventional perturbative
probabilities. In the second step, the partons are hadronized according 
to the string model, from which the production
vertices of hadrons can be extracted. An event weight is given by
\begin{equation}
W_{\mathrm{BE}} = 1 + \sum_{\mathrm{all~pairs}} 
\cos (\Delta x \cdot \Delta p) \;
\theta \left( \frac{\pi}{2} - | \Delta x \cdot \Delta p | \right) ~,
\end{equation}    
where the cosine factor comes from wave function symmetrization and the 
$\theta$ step function ensures that only small $\Delta x \cdot \Delta p$
contribute. Also three-particle correlations are included in a similar 
spirit. Only primary $\pi$, K, $\rho$ and $\omega$ particles, produced 
directly from the string, are included in the global weight.  The number 
of primary particles of each species being rather small --- e.g. about 
16\% of the charged pions are directly produced --- the weight fluctuations 
are manageably small. The second step is iterated, i.e. the same parton
configuration is re-hadronized, until the weighting procedure gives
acceptance. This reweighting does shift the multiplicities
of produced particles, but rather modestly. Particles from resonances 
(including short-lived ones like the $\rho$) are not part of the global 
weight. Instead, in the third step, decay kinematics is selected according 
to a probability distribution that follows the correlation function. 

Kartvelishvili, Kvatadze and M{\o}ller have studied several models
\cite{Rasmus}. The most extreme is a global weight
\begin{equation}
W_{\mathrm{BE}} = \prod_{\mathrm{all~pairs}} 
\left\{ 1 + \lambda \exp(-Q^2 R^2) \right\} ~,
\label{rasmusweight}
\end{equation}    
which then gives an average weight much above unity, an increased
average multiplicity (that can be tuned away), a much increased
three-jet fraction and a reduced fraction of 
$\Z^0 \to \mathrm{b} \overline{\mathrm{b}}$
decays. Since this is unacceptable, different rescaling schemes for 
the global weights are introduced. One is based on a suppression 
by a constant factor for each pair, another on normalizing to a weight
also involving pairs of non-identical particles. Alternatively
the pair weight in eq.~(\ref{rasmusweight}) is modified to
$1 + \cos(\xi Q R) / \cosh(QR)$ with $\xi = 1.15$. These modifications
reduce the problems noted above but do not solve them; additionally
the rescalings are completely \textit{ad hoc} and are given no physics 
explanation.  

The model of Jadach and Zalewski \cite{Jadach} is based on a subdivision 
of the event into clusters of identical particles, to which a particle
can belong only if it has a neighbour within a distance $Q<0.2$~GeV.
This cut is very visible in the final BE distribution, but is probably
required to keep the clusters of tractable size.
A weight, always above unity, is defined for each cluster, and a
global event weight by the product of cluster weights. Since the
multiplicity is increased by the reweighting, the weights are
rescaled by a factor raised to the total pion multiplicity to bring 
the average multiplicity back. A further common factor is needed
to bring the weights to an average of unity. Also the jet
multiplicity then comes out about right, but issues such as the
flavour composition in $\Z^0$ decays have not been studied.
The average multiplicity of a W pair is about 4\% higher than
the sum of two separate W's.  

Fia{\l}kowski and Wit employ a global weight that contains a sum of 
all possible permutations among identical particles. To retain a 
tractable number of terms to evaluate, the procedure is cut short
at permutations involving at most five particles. Studies with cuts
at lower values indicate that the procedure, at least for the inclusive
BE distribution, should have converged by then. Weights are always above
unity and tend to push up the multiplicity distribution. As above,
a factor raised to the total pion multiplicity is used to restore
the average multiplicity and another common factor applied to produce
correct average weight. The possibilities of a change in the flavour
composition of $\Z^0$ decays or of the jet multiplicity have not been
studied.

Several other algorithms based on global weights have also been proposed 
or studied recently \cite{otherglobal}. Since these other models have 
not been used to study the issue of \tmw4j, and do not offer 
any unique insights in the interpretation of nonunit average global 
weights, we will not comment on them here.
   
\subsection{The W mass determination}

At LEP~2 the average space--time separation between the two W decays
is less than 0.1 fm \cite{TSVK}, to be compared with a typical BE radius 
of around 0.5 fm. When the W's decay to $\q\qbar$ pairs, the quarks
fly apart and stretch strings between themselves. These strings will
overlap in the central region, whereas the outer parts will not
in general. Only in the case that two partons from different W's 
travel out in almost the same direction does the overlap spread also 
to the outer regions, but most such events would not survive standard
selection criteria, used to separate W pair events from backgrounds
such as QCD 3-jets. 

Any BE effects caused by the overlap between the $\W^+$ and $\W^-$ 
hadronization systems should therefore predominantly
occur among the centrally produced, low-momentum particles. In this
region it may not be possible to speak about separate $\W^+$ and
$\W^-$ sources of particle production, but only about one single common 
source. Since the hadrons do not emerge tagged with their origin,
the mass definition has to be based on an experimental clustering
procedure, usually first into four jets and thereafter those paired 
to the two W's \cite{workshopWmass}. Possible biases in the detector 
and the procedure can 
be controlled by studying Monte Carlo events generated with the
$\W^+$ and $\W^-$ hadronization processes decoupled from each other.
The shift in the outcome of the procedure when BE effects are
included in full is then what we loosely refer to as a `W mass shift'.
This does not have to imply that the masses of the W propagators 
in the perturbative graphs are affected. Rather, the 
main point is that our limited understanding of the BE 
phenomenon reduces the ability to `unfold' the hadronic data to
arrive at the partonic picture. 

In our standard local scenario \cite{BE95} we found that a mass shift
of around or even somewhat above 100 MeV could not be excluded.
On the scale of the desired experimental accuracy of maybe 30 MeV
\cite{workshopWmass}, as required for precision tests of the standard 
model, this is a large number. However, put in the context of QCD 
physics in general, the uncertainty is not exceptional, neither on 
an absolute nor on a relative scale. Specifically, for effects related
to nonperturbative physics, uncertainties of the order of a pion mass 
or of $\Lambda_{\mathrm{QCD}}$ are fairly common.
We also found that the assumed `attractive' form of the BE
factor defined in eq.\ (\ref{ftwo}) leads to an enhancement of 
production in the low-momentum region of large overlap between the
$\W^+$ and $\W^-$ sources, at the expense of somewhat faster particles. 
The result is that the W mass shift tends to be positive.

This kind of mass shift does not have to be unique for the momentum 
shift method used in our local approach, but could well arise also in
global weight schemes. Just like in local algorithms, the outcome 
would depend on model details. 

First of all, the BE phenomenon could affect the interpretation 
of the W propagators. To see this, it is convenient to start
out from the QED case. The lowest-order process
$\e^+\e^- \to \W^+\W^- \to \ell^+ \nu_{\ell} \ell'^- \nu'_{\ell}$
contains two W masses that are perfectly defined by the momenta of 
the final leptons and neutrinos. If a photon is added to the final 
state, however, there are six charged particles that could have 
radiated it, including all possible interference contributions.
The normal experimental procedure would be either to remove the photon
altogether (relevent for initial-state radiation) or to add it to one 
of the $\W^+$ and $\W^-$ systems. Clearly this is too coarse an 
approximation, in particular for photons well away from the collinear 
regions. So we lose the concept of a unique theoretical or 
experimental definition of the W masses of a given 
event. For the totally inclusive $\W^+\W^-$ cross section there is a 
general proof \cite{proofQED} that the radiative interconnection 
effects are suppressed by $O(\alpha_{\mathrm{em}}\Gamma_{\W}/m_{\W})$. 
The only exception is the Coulomb interaction between two slowly 
moving W's. By contrast, differential distributions could be 
distorted on the level of $O(\alpha_{\mathrm{em}})$. Only in the
limit of vanishing W width would one expect to recover a unique
theoretical separation of radiation. In QED it is always possible 
in principle to calculate the corrections necessary to extract the 
proper average W mass from a given experimental procedure. Since 
complete calculations have not been performed, however, some 
uncertainty may still remain \cite{testQED}.  
   
For QCD there is no radiation from the initial state or the W's 
themselves, but only from the final quarks. Furthermore, colour 
conservation ensures that there are no interconnection effects to
$O(\alpha_{\mathrm{s}})$. The totally inclusive $\W^+\W^-$ cross 
section is therefore protected to  
$O(\alpha_{\mathrm{s}}^2\Gamma_{\W}/m_{\W})$ \cite{proofQED}.
Again differential distributions could contain larger effects,
related to the inability to assign a gluon uniquely to either
of the $\W^+$ and $\W^-$ systems. This perturbative interconnection
is suppressed by propagator effects for energetic gluons, as shown 
in \cite{TSVK}. In the soft region, where gluon energies are below 
the $\Gamma_{\W}$ scale, the propagator damping is not effective,
and non-negligible effects cannot be excluded.
 
Extrapolating from this, it is not impossible that BE effects indeed
have repercussions on the W propagator description. 
To the extent one could still speak
about two different sources of particle production, an effect to a 
global weight would come e.g.\ from interchanging the production of
two identical particles. That is, either pion no.~1 is produced by 
the $\W^+$ and pion no.~2 by the $\W^-$, or the other way around.
Since the two pions have different momenta, in this case one would 
actually be considering interference between Feynman graphs with 
different W propagator masses. Each graph would have to be weighted
with the respective perturbative production matrix elements, in
addition to the BE weight. The exchange of two particles of widely
different momenta is likely to push some W propagator off the mass 
shell and so suppress interference terms. For pairs within the BE 
enhancement region, however, the mass shifts will occur at a scale 
of a few hundred MeV, where the W propagator weight does not vary 
so drastically. The propagator effects are thus not expected to
change the picture dramatically, but could well give some shift of 
the W mass. Since, to the best of our knowledge, none of the global 
models include the W propagators in their weights, this has not been 
put to a quantitative test. Furthermore the hadronization amplitude 
should be complex, cf.\ eq.~(\ref{MBEbomarkus}), as are the W 
propagators, something which could further complicate the 
interference pattern.  

\begin{figure}
  \begin{center}
\setlength{\unitlength}{0.1bp}
\special{!
/gnudict 40 dict def
gnudict begin
/Color false def
/Solid false def
/gnulinewidth 5.000 def
/vshift -33 def
/dl {10 mul} def
/hpt 31.5 def
/vpt 31.5 def
/M {moveto} bind def
/L {lineto} bind def
/R {rmoveto} bind def
/V {rlineto} bind def
/vpt2 vpt 2 mul def
/hpt2 hpt 2 mul def
/Lshow { currentpoint stroke M
  0 vshift R show } def
/Rshow { currentpoint stroke M
  dup stringwidth pop neg vshift R show } def
/Cshow { currentpoint stroke M
  dup stringwidth pop -2 div vshift R show } def
/DL { Color {setrgbcolor Solid {pop []} if 0 setdash }
 {pop pop pop Solid {pop []} if 0 setdash} ifelse } def
/BL { stroke gnulinewidth 2 mul setlinewidth } def
/AL { stroke gnulinewidth 2 div setlinewidth } def
/PL { stroke gnulinewidth setlinewidth } def
/LTb { BL [] 0 0 0 DL } def
/LTa { AL [1 dl 2 dl] 0 setdash 0 0 0 setrgbcolor } def
/LT0 { PL [] 0 1 0 DL } def
/LT1 { PL [4 dl 2 dl] 0 0 1 DL } def
/LT2 { PL [2 dl 3 dl] 1 0 0 DL } def
/LT3 { PL [1 dl 1.5 dl] 1 0 1 DL } def
/LT4 { PL [5 dl 2 dl 1 dl 2 dl] 0 1 1 DL } def
/LT5 { PL [4 dl 3 dl 1 dl 3 dl] 1 1 0 DL } def
/LT6 { PL [2 dl 2 dl 2 dl 4 dl] 0 0 0 DL } def
/LT7 { PL [2 dl 2 dl 2 dl 2 dl 2 dl 4 dl] 1 0.3 0 DL } def
/LT8 { PL [2 dl 2 dl 2 dl 2 dl 2 dl 2 dl 2 dl 4 dl] 0.5 0.5 0.5 DL } def
/P { stroke [] 0 setdash
  currentlinewidth 2 div sub M
  0 currentlinewidth V stroke } def
/D { stroke [] 0 setdash 2 copy vpt add M
  hpt neg vpt neg V hpt vpt neg V
  hpt vpt V hpt neg vpt V closepath stroke
  P } def
/A { stroke [] 0 setdash vpt sub M 0 vpt2 V
  currentpoint stroke M
  hpt neg vpt neg R hpt2 0 V stroke
  } def
/B { stroke [] 0 setdash 2 copy exch hpt sub exch vpt add M
  0 vpt2 neg V hpt2 0 V 0 vpt2 V
  hpt2 neg 0 V closepath stroke
  P } def
/C { stroke [] 0 setdash exch hpt sub exch vpt add M
  hpt2 vpt2 neg V currentpoint stroke M
  hpt2 neg 0 R hpt2 vpt2 V stroke } def
/T { stroke [] 0 setdash 2 copy vpt 1.12 mul add M
  hpt neg vpt -1.62 mul V
  hpt 2 mul 0 V
  hpt neg vpt 1.62 mul V closepath stroke
  P  } def
/S { 2 copy A C} def
end
}
\begin{picture}(2880,1728)(0,0)
\special{"
gnudict begin
gsave
50 50 translate
0.100 0.100 scale
0 setgray
/Helvetica findfont 100 scalefont setfont
newpath
-500.000000 -500.000000 translate
LTa
600 251 M
2097 0 V
600 251 M
0 1426 V
LTb
600 251 M
63 0 V
2034 0 R
-63 0 V
600 568 M
63 0 V
2034 0 R
-63 0 V
600 885 M
63 0 V
2034 0 R
-63 0 V
600 1202 M
63 0 V
2034 0 R
-63 0 V
600 1519 M
63 0 V
2034 0 R
-63 0 V
600 251 M
0 63 V
0 1363 R
0 -63 V
1019 251 M
0 63 V
0 1363 R
0 -63 V
1439 251 M
0 63 V
0 1363 R
0 -63 V
1858 251 M
0 63 V
0 1363 R
0 -63 V
2278 251 M
0 63 V
0 1363 R
0 -63 V
2697 251 M
0 63 V
0 1363 R
0 -63 V
600 251 M
2097 0 V
0 1426 V
-2097 0 V
600 251 L
LT0
2394 1514 M
180 0 V
626 328 M
53 364 V
52 412 V
52 279 V
53 151 V
52 46 V
53 -16 V
52 -51 V
53 -71 V
52 -70 V
52 -77 V
53 -77 V
52 -76 V
53 -66 V
52 -67 V
53 -58 V
52 -49 V
52 -49 V
53 -49 V
52 -40 V
53 -36 V
52 -36 V
53 -31 V
52 -29 V
52 -28 V
53 -27 V
52 -20 V
53 -22 V
52 -18 V
53 -19 V
52 -15 V
52 -13 V
53 -15 V
52 -12 V
53 -13 V
52 -11 V
53 -11 V
52 -8 V
52 -8 V
53 -10 V
LT1
2394 1414 M
180 0 V
626 327 M
53 345 V
52 397 V
52 277 V
53 146 V
52 54 V
53 -7 V
52 -46 V
53 -63 V
52 -74 V
52 -71 V
53 -77 V
52 -70 V
53 -63 V
52 -62 V
53 -59 V
52 -50 V
52 -51 V
53 -46 V
52 -40 V
53 -37 V
52 -35 V
53 -30 V
52 -28 V
52 -26 V
53 -28 V
52 -21 V
53 -22 V
52 -16 V
53 -20 V
52 -15 V
52 -16 V
53 -12 V
52 -14 V
53 -11 V
52 -12 V
53 -11 V
52 -10 V
52 -7 V
53 -9 V
stroke
grestore
end
showpage
}
\put(2334,1414){\makebox(0,0)[r]{{\scriptsize $m_{W^+}+m_{W_-}<2\langle m_W\rangle$}}}
\put(2334,1514){\makebox(0,0)[r]{{\scriptsize $m_{W^+}+m_{W_-}>2\langle m_W\rangle$}}}
\put(1648,51){\makebox(0,0){$Q$ (GeV)}}
\put(100,964){%
\special{ps: gsave currentpoint currentpoint translate
270 rotate neg exch neg exch translate}%
\makebox(0,0)[b]{\shortstack{$1/N_{\mbox{\scriptsize ev}} dN_{\mbox{\scriptsize pair}}/dQ$}}%
\special{ps: currentpoint grestore moveto}%
}
\put(2697,151){\makebox(0,0){5}}
\put(2278,151){\makebox(0,0){4}}
\put(1858,151){\makebox(0,0){3}}
\put(1439,151){\makebox(0,0){2}}
\put(1019,151){\makebox(0,0){1}}
\put(600,151){\makebox(0,0){0}}
\put(540,1519){\makebox(0,0)[r]{0.4}}
\put(540,1202){\makebox(0,0)[r]{0.3}}
\put(540,885){\makebox(0,0)[r]{0.2}}
\put(540,568){\makebox(0,0)[r]{0.1}}
\put(540,251){\makebox(0,0)[r]{0}}
\end{picture}
    \caption[dummy]{{\it The correlation function for pairs of pions
        with one pion from each W as a function of $Q$ for two samples
        of \ee2ww\ events at 170 GeV center of
        mass energy. The full (dashed) line corresponds to events
        where the average mass of the two W's is above (below) the
        nominal W-mass. Both curves are normalized to unity.}}
    \label{fig:hilo}
  \end{center}
\end{figure}
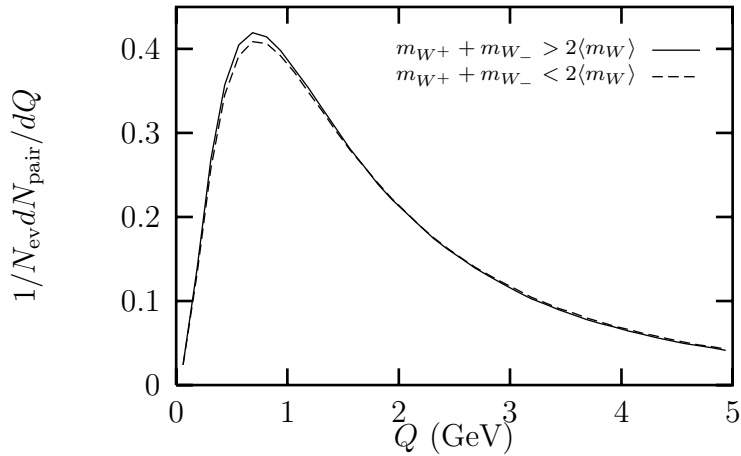

Another way a mass shift could arise in a global weight model is due
to the fact that, for a given total energy, a heavy W will be less
boosted away from the interaction point than a light one. This means
that, for events with high-mass W's, the two fragmentation regions will
have a larger overlap. A pair with one pion from each W is then more 
likely to be close to each other than in events with light-mass W's,
as shown in Fig.~\ref{fig:hilo}. Events with heavier W's would thus 
be given a higher weight (provided the BE weight factor is always 
above unity), which could introduce a mass shift. Also, for a global
weight model that does not conserve multiplicity, one would expect a
higher weight for events with heavier W's, since the multiplicity
increases with the mass.

In more complicated models, with a single source of particle production, 
the W mass concept would be questioned from the onset. However, we do
not really know how to formulate such models, so all the ones studied
to date are based on having a picture with two separate W's as
starting point. 

In the studies of Andersson and Ringn\'er the separation is an
essential part of the model. The matrix element and weight
expressions, eqs.\ (\ref{MBEbomarkus}) and (\ref{WBEbomarkus}),
respectively, are based on a definition of the area spanned by each
string. Therefore the weight of a pair of strings is the product of
the weight of the respective string. If weights are rescaled to unity
average for a string of any mass, it then follows by definition that
the W mass is unaffected.  It has also been shown \cite{JariMarkus}
that effects are negligibly small, below $\sim 10$~MeV, even when the 
weights are not rescaled. In this case a mass shift in principle could 
come from the variation of the average BE weight with the W mass, so the
nonobservation of an effect can be reinterpreted in weight terms, but
we remind that $\Z^0$ and other data in principle exclude this use of
nonunit average weights.

One should here recall the UA1 and E735 studies \cite{UAonedata}, which 
showed a decreasing $\lambda$ parameter with increasing multiplicity
density. This would arise quite naturally if large multiplicities
were a consequence of having many strings in an event \cite{multint},
with no BE cross-talk between strings. The simultaneous observation
of an increasing BE radius $R$ could be used to argue {\em for} 
the existence of cross-talk, however, so it may be premature to use 
UA1/E735 data as argument against a W mass shift.

The studies of Todorova--Nov\'a and Rame\v{s} \cite{Sharka} also give a 
null result, within the statistical uncertainty of $\sim 10$ MeV. This 
holds both for the average mass and a fitted mass peak value. Like in the
previous model, the primary particle production factorizes into two
sources by default. The `theory' classification of particles into two 
groups would then still give unchanged masses. Several alternative 
scenarios were tried, checking for effects coming from misassigned 
particles and from a possible breaking of factorization, but none of 
them gives significant effects.

Kartvelishvili, Kvatadze and M{\o}ller do find a W mass shift with their
methods \cite{Rasmus}, where the BE weight of an event is truly global,
i.e.\ is not just the product of two separate weights but also contains
cross-terms with one particle in a pair from each W. The shift in the 
average mass ranges between 20 and 75 MeV at 175 GeV and between
34 and 92 MeV at 192 GeV for the models studied. However, the authors
note that the use of an average mass shift may be partly misleading,
since typical experimental procedures are based on a fit to a central
mass peak, so that the wings of the Breit-Wigners are suppressed in 
relative importance compared with a straight averaging. Within such
a fitting procedure, the mass shift is still there but never larger 
than about 15 MeV, i.e.\ on an acceptable level.

Jadach and Zalewski, on the other hand, do not find a significant mass 
shift at all \cite{Jadach}: any possible signal is below the statistical
error of 12 MeV. Again this is based on a fit to the mass peak. The
model is reminiscent of one alternative studied by the previous authors,
but uses a BE radius $R$ of 1 fm rather than the 0.5 fm used there.
Since the BE-affected phase space volume is reduced by an increased 
$R$, and since the cut $Q<0.2$~GeV gives a further reduction, there does
not appear to be any contradiction between these two studies 
\cite{Rasmus}.

Also Fia{\l}kowski and Wit fail to find a significant mass shift, and
quote a limit of 20~MeV \cite{fial}. Their Fig.~2 shows a very notable
change of the shape of the W mass spectrum, however. The peak rate is 
reduced, while the rate in the wings is increased. This may indicate 
that the weight rescaling procedure is too simpleminded. 

Even with the wings removed, the fitted W width is increased by 58 MeV
when BE effects are included \cite{fialpriv}. Since the fitting error
is of the order of 30 MeV, the result would seem barely statistically
significant. However, a visual inspection of their Fig.~2 leaves
little doubt that the peak is broadened by BE, so the qualitative
picture is not in question even if the exact number may be. If this
broadening is another manifestation of weight rescaling imperfections
then any results on the average W mass can hardly be trusted. If, on
the other hand, it is a genuine consequence of the model, then it is
in itself an even more interesting phenomenon than a shift of the peak
position, and much simpler to study experimentally. Also the studies
of Jadach and Zalewski give a fitted W width that increases with the
inclusion of BE effects, by about the same amount as above
\cite{Jadach}. Here, however, it is less easy to see from the curves
in the paper whether this is a real phenomenon or just a fluke of the
fitting procedure. For the other global models we have no information.
More studies by the respective authors are here certainly called for,
and below we report on results for our models.

In summary, we thus see that there is no unique answer. Many null 
results have been obtained, but also some nonzero ones. Some of the
models may change the measurable W width even if the average W mass
is unaffected. Obviously,
to claim that the problem has `gone away', it is not enough to find
{\em one} method that give negligible mass or width shifts: one must find
some reason to exclude {\em every} model that give uncomfortable
values.  We are not there yet. However, some of the criticism of 
our original study should be taken seriously, and below we study
a few possible improvements.

\section{New local algorithms}
\label{sec:alg}

Probably the largest weakness of our local approach is the issue how
to conserve the total four-momentum. The procedure described in section
\ref{sec:local} preserves three-momentum locally, but at the expense 
of not conserving energy. The subsequent rescaling of all momenta
by a common factor (in the rest frame of the event) to restore
energy conservation is purely {\it ad hoc}. For studies of a single
$\Z^0$ decay, it can plausibly be argued that such a rescaling does
minimal harm. The same need not hold for a pair of resonances.
Indeed, studies \cite{BE95} show that this global rescaling scheme, 
which we will denote $\BE_0$, introduces an artificial negative shift 
in \tmw4j, making it difficult (although doable) to study the true 
BE effects in this case. This is one reason to consider alternatives.

The global rescaling is also running counter to our original
starting point that BE effects should be local. To be more specific,
we assume that the energy density of the string is a fixed quantity.
To the extent that a pair of particles have their four-momenta slightly 
shifted, the string should act as a `commuting vessel', providing the
difference to other particles produced in the same local region of 
the string. What this means in reality is still not completely  
specified, so further assumptions are necessary. In the
following we discuss four possible algorithms, whereof the last two
are based strictly on the local conservation aspect above, while the 
first two are attempting a slightly different twist to the locality
concept. All are based on
calculating an additional shift $\delta\bfr_k^l$ for some pairs of
particles, where particles $k$ and $l$ need not be identical bosons.
In the end each particle momentum will then be shifted to $\bfp_i' =
\bfp_i + \sum_{j \neq i} \delta \bfp_i^j + \alpha\sum_{k \neq i}
\delta \bfr_i^k$, with the parameter $\alpha$ adjusted separately for 
each event so that the total energy is conserved. 

In the first approach we emulate the criticism of the global
event weight methods with weights always above unity, as being 
intrinsically unstable. It appears more plausible that weights
fluctuate above and below unity. For instance, the simple pair 
symmetrization weight is $1 + \cos (\Delta x \cdot \Delta p)$,
with the $1 + \lambda \exp(-Q^2R^2)$ form only obtained after 
integration over a Gaussian source. Non-Gaussian sources give
oscillatory behaviours, e.g.\ the conventional 
Kopylov--Podgoretski\u{\i} parametrization
for particle production from a spherical surface \cite{KP}.
The global model by Andersson, Hofmann and Ringn\'er is an example 
of weights above as well as below unity. In this case the 
oscillations contain the $\cos (\Delta x \cdot \Delta p)$ 
behaviour dampened by further factors at large values. 

If weights above unity correspond to a shift of pairs towards
smaller relative $Q$ values, the below-unity weights instead give
a shift towards larger $Q$. One therefore is lead to a picture 
where very nearby identical particles are shifted closer, 
those somewhat further are shifted apart, those even further yet
again shifted closer, and so on. Probably the oscillations
dampen out rather quickly, as indicated both by data and by 
the global model studies. We therefore simplify by simulating
only the first peak and dip. Furthermore, to include the desired
damping and to make contact with our normal generation algorithm
(for simplicity), we retain the Gaussian form, but the standard 
$f_2(Q) = 1 + \lambda \exp(-Q^2R^2)$ is multiplied by a further
factor $1 + \alpha \lambda \exp(-Q^2R^2/9)$. The factor $1/9$ in the 
exponential, i.e.\ a factor 3 difference in the $Q$ variable,
is consistent with data and also with what one might expect from 
a dampened $\cos$ form, but should be viewed more as a simple
ansatz than having any deep meaning. 

In the  algorithm, which we denote $\BE_3$, $\delta\bfr_i^j$ is then
non-zero only for pairs of identical bosons, and is calculated in 
the same way as $\delta\bfp_i^j$, with the additional factor $1/9$ in 
the exponential. As explained above, the $\delta\bfr_i^j$ shifts
are then scaled by a common factor $\alpha$ that ensures total energy
conservation. It turns out that the average $\alpha$ needed is 
$\approx -0.2$. The negative sign is exactly what we want to
ensure that $\delta\bfr_i^j$ corresponds to shifting a pair apart, 
while the order of $\alpha$ is consistent with the expected increase 
in the number of affected pairs when a smaller effective radius $R/3$ 
is used. One shortcoming of the method, as implemented here, is
that the input $f_2(0)$ is not quite 2 for $\lambda = 1$ but rather 
$(1 + \lambda) (1 + \alpha \lambda) \approx 1.6$. This could be
solved by starting off with an input $\lambda$ somewhat above unity.

The second algorithm, denoted $\BE_{23}$, is a modification of  
the $\BE_3$ form intended to give $C_2(0) = 1 + \lambda$. The ansatz
is
\begin{equation}
 f_2(Q) = \left\{ 1 + \lambda \exp(-Q^2R^2) \right\}
 \left\{ 1 + \alpha \lambda \exp(-Q^2R^2/9) 
 \left( 1 - \exp(-Q^2R^2/4) \right) \right\} ~,
\end{equation}
which is again applied only to identical pairs. The combination
$\exp(-Q^2R^2/9) \left( 1 - \exp(-Q^2R^2/4) \right)$ can be
viewed as a Gaussian, smeared-out representation of the first
dip of the $\cos$ function. As a technical trick, the 
$\delta\bfr_i^j$ are found as in the $\BE_3$ algorithm and 
thereafter scaled down by the $1 - \exp(-Q^2R^2/4)$ factor.
(This procedure does not quite reproduce the formalism of
eq.~(\ref{qshift}), but comes sufficiently close for our 
purpose, given that the ansatz form in itself is somewhat
arbitarary.)  One should note that, even with the above improvement
relative to the $\BE_3$ scheme, the observable two-particle correlation
is lower at small $Q$ than in the $\BE_0$ algorithm, so some 
further tuning of $\lambda$ could be required. In this scheme, 
$\langle \alpha \rangle \approx -0.25$.

It is interesting to note that the `tuning' of $\alpha$ for
energy conservation could have its analogue in global event
weight algorithms. As we have noted above, a global weight 
would have to have an average value of unity to agree with theory
and data, and this could be achieved (brute-force) by tuning the
form of the weight expression appropriately. While our $\alpha$
is tuned event by event, the corresponding shape parameter(s) in 
global weight schemes would be tuned separately for each partonic
configuration. To the extent that global weights start out
close to an average of unity, the required tuning would be rather 
modest. 

In the other two schemes, the original form of $f_2(Q)$ is retained,
and the energy is instead conserved by picking another pair of particles 
that are shifted apart appropriately. That is, for each pair of identical
particles $i$ and $j$, a pair of non-identical particles, $k$ and $l$,
neither identical to $i$ or $j$, is found in the neighborhood of $i$
and $j$. For each shift $\delta\bfp_i^j$, a corresponding
$\delta\bfr_k^l$ is found so that the total energy and momentum in the
$i,j,k,l$ system is conserved. However, the actual momentum shift of 
a particle is formed as the composant of many contributions, so the 
above pair compensation mechanism is not perfect. The mismatch is 
reflected in a nonunit value $\alpha$ used to rescale the 
$\delta\bfr_k^l$ terms. 

The $k,l$ pair should be the particles `closest' to the pair affected 
by the BE shift, in the spirit of local energy conservation. One option 
would here have been to `look behind the scenes' and use information
on the order of production along the string. However, once decays of
short-lived particles are included, such an approach would still
need arbitrary further rules. We therefore stay with the simplifying 
principle of only using the produced particles.

Looking at $\W^+\W^-$ events and a pair $i,j$ with both particles from
the same W, it is not obvious whether the pair $k,l$ should also be
selected only from this W or if all possible pairs should be
considered.  Below we have chosen the latter as default behaviour, but
the former alternative is also studied below.

One obvious measure of closeness is small invariant mass. A first 
choice would then be to pick the combination that minimizes the
invariant mass $m_{ijkl}$ of all four particles. However, such a 
procedure does not reproduce the input $f_2(Q)$ shape very well: both 
the peak height and peak width are significantly reduced, compared with 
what happens in the $\BE_0$ algorithm. The main reason is that either
of $k$ or $l$ may have particles identical to itself in its local
neighbourhood. The momentum compensation shift of $k$ is at random, 
more or less, and therefore tends to smear the BE signal that could
be introduced relative to $k$'s identical partner. Note that, 
if $k$ and its partner are very close in $Q$ to start with, the 
relative change $\delta Q$ required to produce a significant BE effect 
is very small, approximately $\delta Q \propto Q$. The momentum 
compensation shift on $k$ can therefore easily become larger than the
BE shift proper. 

It is therefore necessary to disfavour momentum compensation shifts
that break up close identical pairs. One alternative would have been
to share the momentum conservation shifts suitably inside such pairs.
We have taken a simpler course, by introducing a suppression factor
$1 - \exp(-Q_k^2 R^2)$ for particle $k$, where $Q_k$ is the $Q$ 
value between $k$ and its nearest identical partner. The form is fixed 
such that a $Q_k = 0$ is forbidden and then the rise matches the
shape of the BE distribution itself. Specifically, in the third 
algorithm, $\BE_m$, the pair $k,l$ is chosen so that the measure
\begin{equation}
  W_{ijkl} = \frac{ (1 - \exp(-Q_k^2 R^2))(1 - \exp(-Q_l^2 R^2))}%
{m^2_{ijkl}}
\end{equation}
is maximized. The average $\alpha$ value required to rescale for the 
effect of multiple shifts is 0.73,  i.e.\ somewhat below unity.

The $\BE_\lambda$ algorithm is inspired by the so-called $\lambda$ measure 
\cite{fractal} (not the be confused with the $\lambda$ parameter of
$f_2(Q)$). It corresponds to a string length in the Lund string 
fragmentation framework. It can be shown that partons in a string are 
colour-connected in a way that tends to minimize this measure. The same is
true for the ordering of the produced hadrons, although with large 
fluctuations. As above, having identical particles nearby to $k,l$ gives
undesirable side effects. Therefore the selection is made so that
\begin{equation}
  W_{ijkl} = \frac{ (1 - \exp(-Q_k^2 R^2))(1 - \exp(-Q_l^2 R^2))}%
{\min_{\mathrm{(12~permutations)}} (m_{ij}m_{jk}m_{kl},%
m_{ij}m_{jl}m_{lk},\ldots)}
\end{equation}
is maximized. The denominator is intended to correspond to $\exp\lambda$.
For cases where particles $i$ and $j$ comes from the
same string, this would favour compensating the energy using particles
that are close by and in the same string. This is thus close in
spirit to some of the global approaches \cite{AndRin,Sharka}. 
We find $\langle\alpha\rangle\approx 0.73$, as above. 

\section{Results}
\label{sec:res}

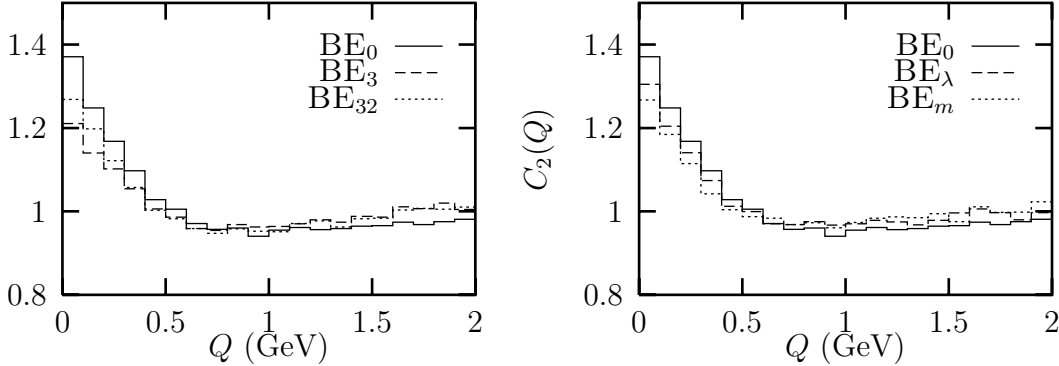
\begin{figure}[t]
   \begin{center}
     \hbox{
       \hskip -1.5cm
\setlength{\unitlength}{0.1bp}
\special{!
/gnudict 40 dict def
gnudict begin
/Color false def
/Solid false def
/gnulinewidth 5.000 def
/vshift -33 def
/dl {10 mul} def
/hpt 31.5 def
/vpt 31.5 def
/M {moveto} bind def
/L {lineto} bind def
/R {rmoveto} bind def
/V {rlineto} bind def
/vpt2 vpt 2 mul def
/hpt2 hpt 2 mul def
/Lshow { currentpoint stroke M
  0 vshift R show } def
/Rshow { currentpoint stroke M
  dup stringwidth pop neg vshift R show } def
/Cshow { currentpoint stroke M
  dup stringwidth pop -2 div vshift R show } def
/DL { Color {setrgbcolor Solid {pop []} if 0 setdash }
 {pop pop pop Solid {pop []} if 0 setdash} ifelse } def
/BL { stroke gnulinewidth 2 mul setlinewidth } def
/AL { stroke gnulinewidth 2 div setlinewidth } def
/PL { stroke gnulinewidth setlinewidth } def
/LTb { BL [] 0 0 0 DL } def
/LTa { AL [1 dl 2 dl] 0 setdash 0 0 0 setrgbcolor } def
/LT0 { PL [] 0 1 0 DL } def
/LT1 { PL [4 dl 2 dl] 0 0 1 DL } def
/LT2 { PL [2 dl 3 dl] 1 0 0 DL } def
/LT3 { PL [1 dl 1.5 dl] 1 0 1 DL } def
/LT4 { PL [5 dl 2 dl 1 dl 2 dl] 0 1 1 DL } def
/LT5 { PL [4 dl 3 dl 1 dl 3 dl] 1 1 0 DL } def
/LT6 { PL [2 dl 2 dl 2 dl 4 dl] 0 0 0 DL } def
/LT7 { PL [2 dl 2 dl 2 dl 2 dl 2 dl 4 dl] 1 0.3 0 DL } def
/LT8 { PL [2 dl 2 dl 2 dl 2 dl 2 dl 2 dl 2 dl 4 dl] 0.5 0.5 0.5 DL } def
/P { stroke [] 0 setdash
  currentlinewidth 2 div sub M
  0 currentlinewidth V stroke } def
/D { stroke [] 0 setdash 2 copy vpt add M
  hpt neg vpt neg V hpt vpt neg V
  hpt vpt V hpt neg vpt V closepath stroke
  P } def
/A { stroke [] 0 setdash vpt sub M 0 vpt2 V
  currentpoint stroke M
  hpt neg vpt neg R hpt2 0 V stroke
  } def
/B { stroke [] 0 setdash 2 copy exch hpt sub exch vpt add M
  0 vpt2 neg V hpt2 0 V 0 vpt2 V
  hpt2 neg 0 V closepath stroke
  P } def
/C { stroke [] 0 setdash exch hpt sub exch vpt add M
  hpt2 vpt2 neg V currentpoint stroke M
  hpt2 neg 0 R hpt2 vpt2 V stroke } def
/T { stroke [] 0 setdash 2 copy vpt 1.12 mul add M
  hpt neg vpt -1.62 mul V
  hpt 2 mul 0 V
  hpt neg vpt 1.62 mul V closepath stroke
  P  } def
/S { 2 copy A C} def
end
}
\begin{picture}(2339,1403)(0,0)
\special{"
gnudict begin
gsave
50 50 translate
0.100 0.100 scale
0 setgray
/Helvetica findfont 100 scalefont setfont
newpath
-500.000000 -500.000000 translate
LTa
600 251 M
0 1101 V
LTb
600 251 M
63 0 V
1493 0 R
-63 0 V
600 566 M
63 0 V
1493 0 R
-63 0 V
600 880 M
63 0 V
1493 0 R
-63 0 V
600 1195 M
63 0 V
1493 0 R
-63 0 V
600 251 M
0 63 V
0 1038 R
0 -63 V
989 251 M
0 63 V
0 1038 R
0 -63 V
1378 251 M
0 63 V
0 1038 R
0 -63 V
1767 251 M
0 63 V
0 1038 R
0 -63 V
2156 251 M
0 63 V
0 1038 R
0 -63 V
600 251 M
1556 0 V
0 1101 V
-1556 0 V
600 251 L
LT0
1853 1189 M
180 0 V
600 1149 M
78 0 V
0 -193 V
78 0 V
0 -126 V
77 0 V
0 -111 V
78 0 V
0 -109 V
78 0 V
0 -36 V
78 0 V
0 -54 V
78 0 V
0 -22 V
77 0 V
0 5 V
78 0 V
0 -31 V
78 0 V
0 23 V
78 0 V
0 10 V
78 0 V
0 -8 V
77 0 V
0 4 V
78 0 V
0 9 V
78 0 V
0 2 V
78 0 V
0 13 V
78 0 V
0 -9 V
77 0 V
0 11 V
78 0 V
0 9 V
78 0 V
LT1
1853 1089 M
180 0 V
600 897 M
78 0 V
0 -111 V
78 0 V
0 -60 V
77 0 V
0 -75 V
78 0 V
0 -76 V
78 0 V
0 -31 V
78 0 V
0 -43 V
78 0 V
0 -7 V
77 0 V
0 22 V
78 0 V
0 -9 V
78 0 V
0 2 V
78 0 V
0 10 V
78 0 V
0 15 V
77 0 V
0 -9 V
78 0 V
0 22 V
78 0 V
0 -3 V
78 0 V
0 39 V
78 0 V
0 -7 V
77 0 V
0 21 V
78 0 V
0 -24 V
78 0 V
LT3
1853 989 M
180 0 V
600 988 M
78 0 V
0 -111 V
78 0 V
0 -120 V
77 0 V
0 -101 V
78 0 V
0 -86 V
78 0 V
0 -32 V
78 0 V
0 -37 V
78 0 V
0 -18 V
77 0 V
0 17 V
78 0 V
0 -9 V
78 0 V
0 -2 V
78 0 V
0 31 V
78 0 V
0 10 V
77 0 V
0 -23 V
78 0 V
0 31 V
78 0 V
0 2 V
78 0 V
0 31 V
78 0 V
0 6 V
77 0 V
0 -3 V
78 0 V
0 8 V
78 0 V
stroke
grestore
end
showpage
}
\put(1793,989){\makebox(0,0)[r]{$\BE_{32}$}}
\put(1793,1089){\makebox(0,0)[r]{$\BE_3$}}
\put(1793,1189){\makebox(0,0)[r]{$\BE_0$}}
\put(1378,51){\makebox(0,0){$Q$ (GeV)}}
\put(100,801){%
\special{ps: gsave currentpoint currentpoint translate
270 rotate neg exch neg exch translate}%
\makebox(0,0)[b]{\shortstack{ }}%
\special{ps: currentpoint grestore moveto}%
}
\put(2156,151){\makebox(0,0){2}}
\put(1767,151){\makebox(0,0){1.5}}
\put(1378,151){\makebox(0,0){1}}
\put(989,151){\makebox(0,0){0.5}}
\put(600,151){\makebox(0,0){0}}
\put(540,1195){\makebox(0,0)[r]{1.4}}
\put(540,880){\makebox(0,0)[r]{1.2}}
\put(540,566){\makebox(0,0)[r]{1}}
\put(540,251){\makebox(0,0)[r]{0.8}}
\end{picture}
       \hskip -1cm
\setlength{\unitlength}{0.1bp}
\special{!
/gnudict 40 dict def
gnudict begin
/Color false def
/Solid false def
/gnulinewidth 5.000 def
/vshift -33 def
/dl {10 mul} def
/hpt 31.5 def
/vpt 31.5 def
/M {moveto} bind def
/L {lineto} bind def
/R {rmoveto} bind def
/V {rlineto} bind def
/vpt2 vpt 2 mul def
/hpt2 hpt 2 mul def
/Lshow { currentpoint stroke M
  0 vshift R show } def
/Rshow { currentpoint stroke M
  dup stringwidth pop neg vshift R show } def
/Cshow { currentpoint stroke M
  dup stringwidth pop -2 div vshift R show } def
/DL { Color {setrgbcolor Solid {pop []} if 0 setdash }
 {pop pop pop Solid {pop []} if 0 setdash} ifelse } def
/BL { stroke gnulinewidth 2 mul setlinewidth } def
/AL { stroke gnulinewidth 2 div setlinewidth } def
/PL { stroke gnulinewidth setlinewidth } def
/LTb { BL [] 0 0 0 DL } def
/LTa { AL [1 dl 2 dl] 0 setdash 0 0 0 setrgbcolor } def
/LT0 { PL [] 0 1 0 DL } def
/LT1 { PL [4 dl 2 dl] 0 0 1 DL } def
/LT2 { PL [2 dl 3 dl] 1 0 0 DL } def
/LT3 { PL [1 dl 1.5 dl] 1 0 1 DL } def
/LT4 { PL [5 dl 2 dl 1 dl 2 dl] 0 1 1 DL } def
/LT5 { PL [4 dl 3 dl 1 dl 3 dl] 1 1 0 DL } def
/LT6 { PL [2 dl 2 dl 2 dl 4 dl] 0 0 0 DL } def
/LT7 { PL [2 dl 2 dl 2 dl 2 dl 2 dl 4 dl] 1 0.3 0 DL } def
/LT8 { PL [2 dl 2 dl 2 dl 2 dl 2 dl 2 dl 2 dl 4 dl] 0.5 0.5 0.5 DL } def
/P { stroke [] 0 setdash
  currentlinewidth 2 div sub M
  0 currentlinewidth V stroke } def
/D { stroke [] 0 setdash 2 copy vpt add M
  hpt neg vpt neg V hpt vpt neg V
  hpt vpt V hpt neg vpt V closepath stroke
  P } def
/A { stroke [] 0 setdash vpt sub M 0 vpt2 V
  currentpoint stroke M
  hpt neg vpt neg R hpt2 0 V stroke
  } def
/B { stroke [] 0 setdash 2 copy exch hpt sub exch vpt add M
  0 vpt2 neg V hpt2 0 V 0 vpt2 V
  hpt2 neg 0 V closepath stroke
  P } def
/C { stroke [] 0 setdash exch hpt sub exch vpt add M
  hpt2 vpt2 neg V currentpoint stroke M
  hpt2 neg 0 R hpt2 vpt2 V stroke } def
/T { stroke [] 0 setdash 2 copy vpt 1.12 mul add M
  hpt neg vpt -1.62 mul V
  hpt 2 mul 0 V
  hpt neg vpt 1.62 mul V closepath stroke
  P  } def
/S { 2 copy A C} def
end
}
\begin{picture}(2339,1403)(0,0)
\special{"
gnudict begin
gsave
50 50 translate
0.100 0.100 scale
0 setgray
/Helvetica findfont 100 scalefont setfont
newpath
-500.000000 -500.000000 translate
LTa
600 251 M
0 1101 V
LTb
600 251 M
63 0 V
1493 0 R
-63 0 V
600 566 M
63 0 V
1493 0 R
-63 0 V
600 880 M
63 0 V
1493 0 R
-63 0 V
600 1195 M
63 0 V
1493 0 R
-63 0 V
600 251 M
0 63 V
0 1038 R
0 -63 V
989 251 M
0 63 V
0 1038 R
0 -63 V
1378 251 M
0 63 V
0 1038 R
0 -63 V
1767 251 M
0 63 V
0 1038 R
0 -63 V
2156 251 M
0 63 V
0 1038 R
0 -63 V
600 251 M
1556 0 V
0 1101 V
-1556 0 V
600 251 L
LT0
1853 1189 M
180 0 V
600 1149 M
78 0 V
0 -193 V
78 0 V
0 -126 V
77 0 V
0 -111 V
78 0 V
0 -109 V
78 0 V
0 -36 V
78 0 V
0 -54 V
78 0 V
0 -22 V
77 0 V
0 5 V
78 0 V
0 -31 V
78 0 V
0 23 V
78 0 V
0 10 V
78 0 V
0 -8 V
77 0 V
0 4 V
78 0 V
0 9 V
78 0 V
0 2 V
78 0 V
0 13 V
78 0 V
0 -9 V
77 0 V
0 11 V
78 0 V
0 9 V
78 0 V
LT1
1853 1089 M
180 0 V
600 1045 M
78 0 V
0 -158 V
78 0 V
0 -100 V
77 0 V
0 -105 V
78 0 V
0 -97 V
78 0 V
0 -20 V
78 0 V
0 -46 V
78 0 V
0 -3 V
77 0 V
0 11 V
78 0 V
0 -13 V
78 0 V
0 5 V
78 0 V
0 13 V
78 0 V
0 -6 V
77 0 V
0 -11 V
78 0 V
0 17 V
78 0 V
0 28 V
78 0 V
0 15 V
78 0 V
0 -14 V
77 0 V
0 -27 V
78 0 V
0 27 V
78 0 V
LT3
1853 989 M
180 0 V
600 986 M
78 0 V
0 -130 V
78 0 V
0 -110 V
77 0 V
0 -114 V
78 0 V
0 -60 V
78 0 V
0 -26 V
78 0 V
0 -6 V
78 0 V
0 -24 V
77 0 V
0 7 V
78 0 V
0 -19 V
78 0 V
0 20 V
78 0 V
0 16 V
78 0 V
0 5 V
77 0 V
0 -3 V
78 0 V
0 15 V
78 0 V
0 -30 V
78 0 V
0 56 V
78 0 V
0 -21 V
77 0 V
0 1 V
78 0 V
0 39 V
78 0 V
stroke
grestore
end
showpage
}
\put(1793,989){\makebox(0,0)[r]{$\BE_m$}}
\put(1793,1089){\makebox(0,0)[r]{$\BE_\lambda$}}
\put(1793,1189){\makebox(0,0)[r]{$\BE_0$}}
\put(1378,51){\makebox(0,0){$Q$ (GeV)}}
\put(280,801){%
\special{ps: gsave currentpoint currentpoint translate
270 rotate neg exch neg exch translate}%
\makebox(0,0)[b]{\shortstack{$C_2(Q)$}}%
\special{ps: currentpoint grestore moveto}%
}
\put(2156,151){\makebox(0,0){2}}
\put(1767,151){\makebox(0,0){1.5}}
\put(1378,151){\makebox(0,0){1}}
\put(989,151){\makebox(0,0){0.5}}
\put(600,151){\makebox(0,0){0}}
\put(540,1195){\makebox(0,0)[r]{1.4}}
\put(540,880){\makebox(0,0)[r]{1.2}}
\put(540,566){\makebox(0,0)[r]{1}}
\put(540,251){\makebox(0,0)[r]{0.8}}
\end{picture}
     }
     \caption[dummy]{{\it The BE enhancement w.r.t.\ the no-BE case of the
         like-signed $\pi\pi$ correlation function in $\Z^0$ decays as a
         function of $Q$.}}
     \label{fig:c2}
  \end{center}
\end{figure}

Armed with these new algorithms we can now proceed to estimate BE
effects. First consider the two-particle correlation function for
like-sign $\pi$ pairs from $\Z^0$ decays normalized to a no-BE world,
Fig.~\ref{fig:c2}.  All four algorithms were used with the same
$\lambda = 1$ and $R = 0.5$~fm, but still show noticable differences.
The enhancement at small $Q$ is smallest in the $\BE_3$ algorithm, as
should be expected from the simpleminded way in which we picked the
form of the energy-compensating below-unity extra factor.  In all
cases we expect that the parameters $\lambda$ and $R$ can be adjusted
to reproduce experimental data.

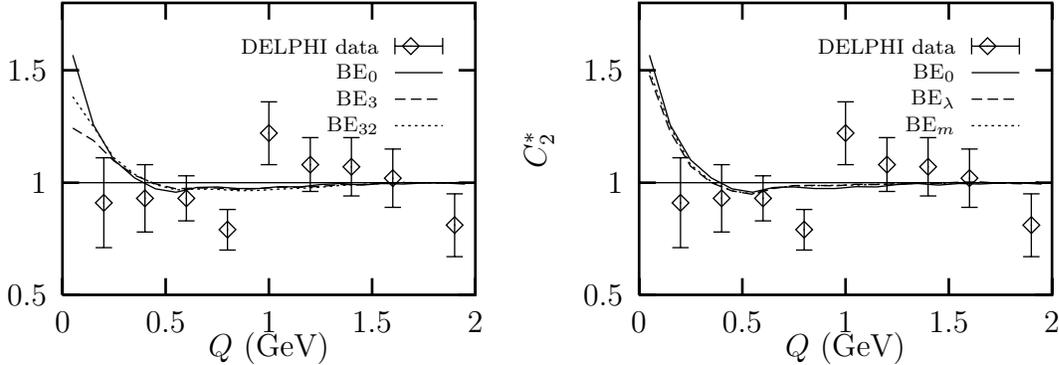
\begin{figure}[t]
  \begin{center}
     \hbox{
       \hskip -1.5cm
\setlength{\unitlength}{0.1bp}
\special{!
/gnudict 40 dict def
gnudict begin
/Color false def
/Solid false def
/gnulinewidth 5.000 def
/vshift -33 def
/dl {10 mul} def
/hpt 31.5 def
/vpt 31.5 def
/M {moveto} bind def
/L {lineto} bind def
/R {rmoveto} bind def
/V {rlineto} bind def
/vpt2 vpt 2 mul def
/hpt2 hpt 2 mul def
/Lshow { currentpoint stroke M
  0 vshift R show } def
/Rshow { currentpoint stroke M
  dup stringwidth pop neg vshift R show } def
/Cshow { currentpoint stroke M
  dup stringwidth pop -2 div vshift R show } def
/DL { Color {setrgbcolor Solid {pop []} if 0 setdash }
 {pop pop pop Solid {pop []} if 0 setdash} ifelse } def
/BL { stroke gnulinewidth 2 mul setlinewidth } def
/AL { stroke gnulinewidth 2 div setlinewidth } def
/PL { stroke gnulinewidth setlinewidth } def
/LTb { BL [] 0 0 0 DL } def
/LTa { AL [1 dl 2 dl] 0 setdash 0 0 0 setrgbcolor } def
/LT0 { PL [] 0 1 0 DL } def
/LT1 { PL [4 dl 2 dl] 0 0 1 DL } def
/LT2 { PL [2 dl 3 dl] 1 0 0 DL } def
/LT3 { PL [1 dl 1.5 dl] 1 0 1 DL } def
/LT4 { PL [5 dl 2 dl 1 dl 2 dl] 0 1 1 DL } def
/LT5 { PL [4 dl 3 dl 1 dl 3 dl] 1 1 0 DL } def
/LT6 { PL [2 dl 2 dl 2 dl 4 dl] 0 0 0 DL } def
/LT7 { PL [2 dl 2 dl 2 dl 2 dl 2 dl 4 dl] 1 0.3 0 DL } def
/LT8 { PL [2 dl 2 dl 2 dl 2 dl 2 dl 2 dl 2 dl 4 dl] 0.5 0.5 0.5 DL } def
/P { stroke [] 0 setdash
  currentlinewidth 2 div sub M
  0 currentlinewidth V stroke } def
/D { stroke [] 0 setdash 2 copy vpt add M
  hpt neg vpt neg V hpt vpt neg V
  hpt vpt V hpt neg vpt V closepath stroke
  P } def
/A { stroke [] 0 setdash vpt sub M 0 vpt2 V
  currentpoint stroke M
  hpt neg vpt neg R hpt2 0 V stroke
  } def
/B { stroke [] 0 setdash 2 copy exch hpt sub exch vpt add M
  0 vpt2 neg V hpt2 0 V 0 vpt2 V
  hpt2 neg 0 V closepath stroke
  P } def
/C { stroke [] 0 setdash exch hpt sub exch vpt add M
  hpt2 vpt2 neg V currentpoint stroke M
  hpt2 neg 0 R hpt2 vpt2 V stroke } def
/T { stroke [] 0 setdash 2 copy vpt 1.12 mul add M
  hpt neg vpt -1.62 mul V
  hpt 2 mul 0 V
  hpt neg vpt 1.62 mul V closepath stroke
  P  } def
/S { 2 copy A C} def
end
}
\begin{picture}(2339,1403)(0,0)
\special{"
gnudict begin
gsave
50 50 translate
0.100 0.100 scale
0 setgray
/Helvetica findfont 100 scalefont setfont
newpath
-500.000000 -500.000000 translate
LTa
600 251 M
0 1101 V
LTb
600 251 M
63 0 V
1493 0 R
-63 0 V
600 674 M
63 0 V
1493 0 R
-63 0 V
600 1098 M
63 0 V
1493 0 R
-63 0 V
600 251 M
0 63 V
0 1038 R
0 -63 V
989 251 M
0 63 V
0 1038 R
0 -63 V
1378 251 M
0 63 V
0 1038 R
0 -63 V
1767 251 M
0 63 V
0 1038 R
0 -63 V
2156 251 M
0 63 V
0 1038 R
0 -63 V
600 251 M
1556 0 V
0 1101 V
-1556 0 V
600 251 L
LT0
1913 1189 D
756 598 D
911 615 D
1067 615 D
1222 497 D
1378 861 D
1534 742 D
1689 734 D
1845 691 D
2078 514 D
1853 1189 M
180 0 V
-180 31 R
0 -62 V
180 62 R
0 -62 V
756 429 M
0 339 V
725 429 M
62 0 V
725 768 M
62 0 V
911 488 M
0 254 V
880 488 M
62 0 V
880 742 M
62 0 V
1067 530 M
0 170 V
1036 530 M
62 0 V
-62 170 R
62 0 V
1222 420 M
0 153 V
1191 420 M
62 0 V
-62 153 R
62 0 V
125 169 R
0 237 V
1347 742 M
62 0 V
-62 237 R
62 0 V
1534 641 M
0 203 V
1503 641 M
62 0 V
-62 203 R
62 0 V
1689 624 M
0 220 V
1658 624 M
62 0 V
-62 220 R
62 0 V
1845 581 M
0 220 V
1814 581 M
62 0 V
-62 220 R
62 0 V
2078 395 M
0 237 V
2047 395 M
62 0 V
-62 237 R
62 0 V
LT0
1853 1089 M
180 0 V
639 1155 M
717 892 L
794 760 L
78 -68 V
78 -41 V
78 -13 V
78 18 V
77 2 V
78 -6 V
78 0 V
78 7 V
78 -1 V
77 9 V
78 4 V
78 -6 V
78 6 V
78 -2 V
77 4 V
78 0 V
78 -4 V
LT1
1853 989 M
180 0 V
639 880 M
78 -47 V
77 -77 V
78 -54 V
78 -32 V
78 -19 V
78 3 V
77 0 V
78 -6 V
78 4 V
78 5 V
78 0 V
77 3 V
78 7 V
78 0 V
78 6 V
78 0 V
77 1 V
78 -1 V
78 0 V
LT3
1853 889 M
180 0 V
639 998 M
717 885 L
794 768 L
78 -62 V
78 -39 V
78 -20 V
78 3 V
77 -1 V
78 -5 V
78 1 V
78 5 V
78 3 V
77 4 V
78 7 V
78 3 V
78 4 V
78 4 V
77 -1 V
78 0 V
78 1 V
LT0
600 674 M
16 0 V
15 0 V
16 0 V
16 0 V
16 0 V
15 0 V
16 0 V
16 0 V
15 0 V
16 0 V
16 0 V
16 0 V
15 0 V
16 0 V
16 0 V
15 0 V
16 0 V
16 0 V
16 0 V
15 0 V
16 0 V
16 0 V
15 0 V
16 0 V
16 0 V
16 0 V
15 0 V
16 0 V
16 0 V
16 0 V
15 0 V
16 0 V
16 0 V
15 0 V
16 0 V
16 0 V
16 0 V
15 0 V
16 0 V
16 0 V
15 0 V
16 0 V
16 0 V
16 0 V
15 0 V
16 0 V
16 0 V
15 0 V
16 0 V
16 0 V
16 0 V
15 0 V
16 0 V
16 0 V
15 0 V
16 0 V
16 0 V
16 0 V
15 0 V
16 0 V
16 0 V
15 0 V
16 0 V
16 0 V
16 0 V
15 0 V
16 0 V
16 0 V
15 0 V
16 0 V
16 0 V
16 0 V
15 0 V
16 0 V
16 0 V
16 0 V
15 0 V
16 0 V
16 0 V
15 0 V
16 0 V
16 0 V
16 0 V
15 0 V
16 0 V
16 0 V
15 0 V
16 0 V
16 0 V
16 0 V
15 0 V
16 0 V
16 0 V
15 0 V
16 0 V
16 0 V
16 0 V
15 0 V
16 0 V
stroke
grestore
end
showpage
}
\put(1793,889){\makebox(0,0)[r]{{\scriptsize $\BE_{32}$}}}
\put(1793,989){\makebox(0,0)[r]{{\scriptsize $\BE_3$}}}
\put(1793,1089){\makebox(0,0)[r]{{\scriptsize $\BE_0$}}}
\put(1793,1189){\makebox(0,0)[r]{{\scriptsize DELPHI data}}}
\put(1378,51){\makebox(0,0){$Q$ (GeV)}}
\put(100,801){%
\special{ps: gsave currentpoint currentpoint translate
270 rotate neg exch neg exch translate}%
\makebox(0,0)[b]{\shortstack{ }}%
\special{ps: currentpoint grestore moveto}%
}
\put(2156,151){\makebox(0,0){2}}
\put(1767,151){\makebox(0,0){1.5}}
\put(1378,151){\makebox(0,0){1}}
\put(989,151){\makebox(0,0){0.5}}
\put(600,151){\makebox(0,0){0}}
\put(540,1098){\makebox(0,0)[r]{1.5}}
\put(540,674){\makebox(0,0)[r]{1}}
\put(540,251){\makebox(0,0)[r]{0.5}}
\end{picture}
       \hskip -1cm
\setlength{\unitlength}{0.1bp}
\special{!
/gnudict 40 dict def
gnudict begin
/Color false def
/Solid false def
/gnulinewidth 5.000 def
/vshift -33 def
/dl {10 mul} def
/hpt 31.5 def
/vpt 31.5 def
/M {moveto} bind def
/L {lineto} bind def
/R {rmoveto} bind def
/V {rlineto} bind def
/vpt2 vpt 2 mul def
/hpt2 hpt 2 mul def
/Lshow { currentpoint stroke M
  0 vshift R show } def
/Rshow { currentpoint stroke M
  dup stringwidth pop neg vshift R show } def
/Cshow { currentpoint stroke M
  dup stringwidth pop -2 div vshift R show } def
/DL { Color {setrgbcolor Solid {pop []} if 0 setdash }
 {pop pop pop Solid {pop []} if 0 setdash} ifelse } def
/BL { stroke gnulinewidth 2 mul setlinewidth } def
/AL { stroke gnulinewidth 2 div setlinewidth } def
/PL { stroke gnulinewidth setlinewidth } def
/LTb { BL [] 0 0 0 DL } def
/LTa { AL [1 dl 2 dl] 0 setdash 0 0 0 setrgbcolor } def
/LT0 { PL [] 0 1 0 DL } def
/LT1 { PL [4 dl 2 dl] 0 0 1 DL } def
/LT2 { PL [2 dl 3 dl] 1 0 0 DL } def
/LT3 { PL [1 dl 1.5 dl] 1 0 1 DL } def
/LT4 { PL [5 dl 2 dl 1 dl 2 dl] 0 1 1 DL } def
/LT5 { PL [4 dl 3 dl 1 dl 3 dl] 1 1 0 DL } def
/LT6 { PL [2 dl 2 dl 2 dl 4 dl] 0 0 0 DL } def
/LT7 { PL [2 dl 2 dl 2 dl 2 dl 2 dl 4 dl] 1 0.3 0 DL } def
/LT8 { PL [2 dl 2 dl 2 dl 2 dl 2 dl 2 dl 2 dl 4 dl] 0.5 0.5 0.5 DL } def
/P { stroke [] 0 setdash
  currentlinewidth 2 div sub M
  0 currentlinewidth V stroke } def
/D { stroke [] 0 setdash 2 copy vpt add M
  hpt neg vpt neg V hpt vpt neg V
  hpt vpt V hpt neg vpt V closepath stroke
  P } def
/A { stroke [] 0 setdash vpt sub M 0 vpt2 V
  currentpoint stroke M
  hpt neg vpt neg R hpt2 0 V stroke
  } def
/B { stroke [] 0 setdash 2 copy exch hpt sub exch vpt add M
  0 vpt2 neg V hpt2 0 V 0 vpt2 V
  hpt2 neg 0 V closepath stroke
  P } def
/C { stroke [] 0 setdash exch hpt sub exch vpt add M
  hpt2 vpt2 neg V currentpoint stroke M
  hpt2 neg 0 R hpt2 vpt2 V stroke } def
/T { stroke [] 0 setdash 2 copy vpt 1.12 mul add M
  hpt neg vpt -1.62 mul V
  hpt 2 mul 0 V
  hpt neg vpt 1.62 mul V closepath stroke
  P  } def
/S { 2 copy A C} def
end
}
\begin{picture}(2339,1403)(0,0)
\special{"
gnudict begin
gsave
50 50 translate
0.100 0.100 scale
0 setgray
/Helvetica findfont 100 scalefont setfont
newpath
-500.000000 -500.000000 translate
LTa
600 251 M
0 1101 V
LTb
600 251 M
63 0 V
1493 0 R
-63 0 V
600 674 M
63 0 V
1493 0 R
-63 0 V
600 1098 M
63 0 V
1493 0 R
-63 0 V
600 251 M
0 63 V
0 1038 R
0 -63 V
989 251 M
0 63 V
0 1038 R
0 -63 V
1378 251 M
0 63 V
0 1038 R
0 -63 V
1767 251 M
0 63 V
0 1038 R
0 -63 V
2156 251 M
0 63 V
0 1038 R
0 -63 V
600 251 M
1556 0 V
0 1101 V
-1556 0 V
600 251 L
LT0
1913 1189 D
756 598 D
911 615 D
1067 615 D
1222 497 D
1378 861 D
1534 742 D
1689 734 D
1845 691 D
2078 514 D
1853 1189 M
180 0 V
-180 31 R
0 -62 V
180 62 R
0 -62 V
756 429 M
0 339 V
725 429 M
62 0 V
725 768 M
62 0 V
911 488 M
0 254 V
880 488 M
62 0 V
880 742 M
62 0 V
1067 530 M
0 170 V
1036 530 M
62 0 V
-62 170 R
62 0 V
1222 420 M
0 153 V
1191 420 M
62 0 V
-62 153 R
62 0 V
125 169 R
0 237 V
1347 742 M
62 0 V
-62 237 R
62 0 V
1534 641 M
0 203 V
1503 641 M
62 0 V
-62 203 R
62 0 V
1689 624 M
0 220 V
1658 624 M
62 0 V
-62 220 R
62 0 V
1845 581 M
0 220 V
1814 581 M
62 0 V
-62 220 R
62 0 V
2078 395 M
0 237 V
2047 395 M
62 0 V
-62 237 R
62 0 V
LT0
1853 1089 M
180 0 V
639 1155 M
717 892 L
794 760 L
78 -68 V
78 -41 V
78 -13 V
78 18 V
77 2 V
78 -6 V
78 0 V
78 7 V
78 -1 V
77 9 V
78 4 V
78 -6 V
78 6 V
78 -2 V
77 4 V
78 0 V
78 -4 V
LT1
1853 989 M
180 0 V
639 1079 M
717 867 L
794 736 L
78 -62 V
78 -31 V
78 -13 V
78 22 V
77 11 V
78 1 V
78 -2 V
78 6 V
78 -1 V
77 3 V
78 2 V
78 0 V
78 4 V
78 1 V
77 -5 V
78 -2 V
78 4 V
LT3
1853 889 M
180 0 V
639 1096 M
717 885 L
794 742 L
78 -64 V
78 -35 V
78 -11 V
78 22 V
77 8 V
78 1 V
78 1 V
78 1 V
78 3 V
77 1 V
78 3 V
78 -1 V
78 3 V
78 -1 V
77 1 V
78 -2 V
78 5 V
LT0
600 674 M
16 0 V
15 0 V
16 0 V
16 0 V
16 0 V
15 0 V
16 0 V
16 0 V
15 0 V
16 0 V
16 0 V
16 0 V
15 0 V
16 0 V
16 0 V
15 0 V
16 0 V
16 0 V
16 0 V
15 0 V
16 0 V
16 0 V
15 0 V
16 0 V
16 0 V
16 0 V
15 0 V
16 0 V
16 0 V
16 0 V
15 0 V
16 0 V
16 0 V
15 0 V
16 0 V
16 0 V
16 0 V
15 0 V
16 0 V
16 0 V
15 0 V
16 0 V
16 0 V
16 0 V
15 0 V
16 0 V
16 0 V
15 0 V
16 0 V
16 0 V
16 0 V
15 0 V
16 0 V
16 0 V
15 0 V
16 0 V
16 0 V
16 0 V
15 0 V
16 0 V
16 0 V
15 0 V
16 0 V
16 0 V
16 0 V
15 0 V
16 0 V
16 0 V
15 0 V
16 0 V
16 0 V
16 0 V
15 0 V
16 0 V
16 0 V
16 0 V
15 0 V
16 0 V
16 0 V
15 0 V
16 0 V
16 0 V
16 0 V
15 0 V
16 0 V
16 0 V
15 0 V
16 0 V
16 0 V
16 0 V
15 0 V
16 0 V
16 0 V
15 0 V
16 0 V
16 0 V
16 0 V
15 0 V
16 0 V
stroke
grestore
end
showpage
}
\put(1793,889){\makebox(0,0)[r]{{\scriptsize $\BE_m$}}}
\put(1793,989){\makebox(0,0)[r]{{\scriptsize $\BE_\lambda$}}}
\put(1793,1089){\makebox(0,0)[r]{{\scriptsize $\BE_0$}}}
\put(1793,1189){\makebox(0,0)[r]{{\scriptsize DELPHI data}}}
\put(1378,51){\makebox(0,0){$Q$ (GeV)}}
\put(280,801){%
\special{ps: gsave currentpoint currentpoint translate
270 rotate neg exch neg exch translate}%
\makebox(0,0)[b]{\shortstack{$C_2^*$}}%
\special{ps: currentpoint grestore moveto}%
}
\put(2156,151){\makebox(0,0){2}}
\put(1767,151){\makebox(0,0){1.5}}
\put(1378,151){\makebox(0,0){1}}
\put(989,151){\makebox(0,0){0.5}}
\put(600,151){\makebox(0,0){0}}
\put(540,1098){\makebox(0,0)[r]{1.5}}
\put(540,674){\makebox(0,0)[r]{1}}
\put(540,251){\makebox(0,0)[r]{0.5}}
\end{picture}
     }
    \caption[dummy]{{\it The ratio between the like-signed and
        unlike-signed $\pi\pi$ correlation function as a function of
        $Q$, restricted to pairs of particles stemming from different W
        bosons in \ee2ww\ events at \lep2\ according to the procedure in
        \cite{NoWWBE}.}}
    \label{fig:c2ww}
  \end{center}
\end{figure}

In the introduction we mentioned the result presented by the DELPHI
collaboration \cite{NoWWBE}, where they found no trace of BE
correlations among particles from different W bosons in fully hadronic
\ee2ww\ event. This was done by studying the ratio
\begin{equation}
C^*_2(Q) = \frac{ N^{\pm\pm}_{\W\W \to 4j}(Q) - 
2 N^{\pm\pm}_{\W\W \to 2j\ell\nu}(Q) }%
{ N^{+-}_{\W\W \to 4j}(Q) - 2 N^{+-}_{\W\W \to 2j\ell\nu}(Q) }~.
\label{Rstarstar}
\end{equation}
Thus the numerator is the distribution in $Q$ of like-sign pairs from 
fully hadronic events, subtracted with twice the distribution from 
semi-leptonic events. In the limit that the two W's hadronize 
completely independently, this difference is then made up of pairs
where one particle comes from each W. The denominator is the same for
unlike-signed pairs, which here should provide a good reference
sample: with one particle of the pair from each W there is not going
to be any of the resonance peaks that appear for distributions inside
a W.  In Fig.~\ref{fig:c2ww} we
compare this result with the prediction from our algorithms, using the
same parameters as in Fig.~\ref{fig:c2}. Contrary to the data our
models predict a clear BE enhancement for $Q$ close to zero. The
experimental statistics (only 24 hadronic and 25 semi-leptonic events
were used) is not large enough to actually rule out the models. During
the lifetime of \lep2, the statistics is expected to grow by a factor
50, by which time it certainly would be possible to rule out our
models, should the absence of BE enhancement in the data persist.

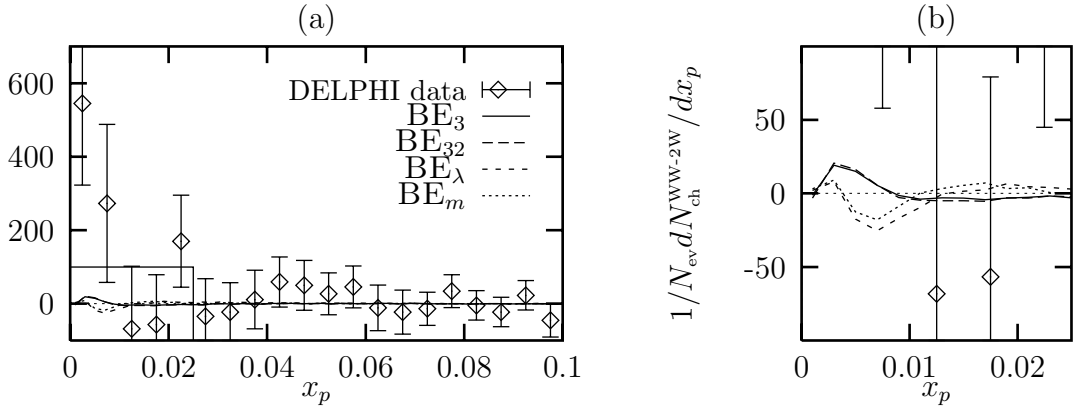
\begin{figure}[t]
  \begin{center}
    \hbox{
    \hskip -1cm
\setlength{\unitlength}{0.1bp}
\special{!
/gnudict 40 dict def
gnudict begin
/Color false def
/Solid false def
/gnulinewidth 5.000 def
/vshift -33 def
/dl {10 mul} def
/hpt 31.5 def
/vpt 31.5 def
/M {moveto} bind def
/L {lineto} bind def
/R {rmoveto} bind def
/V {rlineto} bind def
/vpt2 vpt 2 mul def
/hpt2 hpt 2 mul def
/Lshow { currentpoint stroke M
  0 vshift R show } def
/Rshow { currentpoint stroke M
  dup stringwidth pop neg vshift R show } def
/Cshow { currentpoint stroke M
  dup stringwidth pop -2 div vshift R show } def
/DL { Color {setrgbcolor Solid {pop []} if 0 setdash }
 {pop pop pop Solid {pop []} if 0 setdash} ifelse } def
/BL { stroke gnulinewidth 2 mul setlinewidth } def
/AL { stroke gnulinewidth 2 div setlinewidth } def
/PL { stroke gnulinewidth setlinewidth } def
/LTb { BL [] 0 0 0 DL } def
/LTa { AL [1 dl 2 dl] 0 setdash 0 0 0 setrgbcolor } def
/LT0 { PL [] 0 1 0 DL } def
/LT1 { PL [4 dl 2 dl] 0 0 1 DL } def
/LT2 { PL [2 dl 3 dl] 1 0 0 DL } def
/LT3 { PL [1 dl 1.5 dl] 1 0 1 DL } def
/LT4 { PL [5 dl 2 dl 1 dl 2 dl] 0 1 1 DL } def
/LT5 { PL [4 dl 3 dl 1 dl 3 dl] 1 1 0 DL } def
/LT6 { PL [2 dl 2 dl 2 dl 4 dl] 0 0 0 DL } def
/LT7 { PL [2 dl 2 dl 2 dl 2 dl 2 dl 4 dl] 1 0.3 0 DL } def
/LT8 { PL [2 dl 2 dl 2 dl 2 dl 2 dl 2 dl 2 dl 4 dl] 0.5 0.5 0.5 DL } def
/P { stroke [] 0 setdash
  currentlinewidth 2 div sub M
  0 currentlinewidth V stroke } def
/D { stroke [] 0 setdash 2 copy vpt add M
  hpt neg vpt neg V hpt vpt neg V
  hpt vpt V hpt neg vpt V closepath stroke
  P } def
/A { stroke [] 0 setdash vpt sub M 0 vpt2 V
  currentpoint stroke M
  hpt neg vpt neg R hpt2 0 V stroke
  } def
/B { stroke [] 0 setdash 2 copy exch hpt sub exch vpt add M
  0 vpt2 neg V hpt2 0 V 0 vpt2 V
  hpt2 neg 0 V closepath stroke
  P } def
/C { stroke [] 0 setdash exch hpt sub exch vpt add M
  hpt2 vpt2 neg V currentpoint stroke M
  hpt2 neg 0 R hpt2 vpt2 V stroke } def
/T { stroke [] 0 setdash 2 copy vpt 1.12 mul add M
  hpt neg vpt -1.62 mul V
  hpt 2 mul 0 V
  hpt neg vpt 1.62 mul V closepath stroke
  P  } def
/S { 2 copy A C} def
end
}
\begin{picture}(2519,1511)(0,0)
\special{"
gnudict begin
gsave
50 50 translate
0.100 0.100 scale
0 setgray
/Helvetica findfont 100 scalefont setfont
newpath
-500.000000 -500.000000 translate
LTa
480 390 M
1856 0 V
480 251 M
0 1109 V
LTb
480 390 M
63 0 V
1793 0 R
-63 0 V
480 667 M
63 0 V
1793 0 R
-63 0 V
480 944 M
63 0 V
1793 0 R
-63 0 V
480 1221 M
63 0 V
1793 0 R
-63 0 V
480 251 M
0 63 V
0 1046 R
0 -63 V
851 251 M
0 63 V
0 1046 R
0 -63 V
1222 251 M
0 63 V
0 1046 R
0 -63 V
1594 251 M
0 63 V
0 1046 R
0 -63 V
1965 251 M
0 63 V
0 1046 R
0 -63 V
2336 251 M
0 63 V
0 1046 R
0 -63 V
480 251 M
1856 0 V
0 1109 V
-1856 0 V
480 251 L
LT0
2093 1197 D
526 1145 D
619 768 D
712 295 D
805 311 D
898 625 D
990 342 D
1083 358 D
1176 405 D
1269 472 D
1362 459 D
1454 427 D
1547 453 D
1640 374 D
1733 358 D
1826 371 D
1918 437 D
2011 383 D
2104 358 D
2197 421 D
2290 327 D
2033 1197 M
180 0 V
-180 31 R
0 -62 V
180 62 R
0 -62 V
526 837 M
0 523 V
495 837 M
62 0 V
-62 523 R
62 0 V
619 470 M
0 596 V
588 470 M
62 0 V
-62 596 R
62 0 V
712 251 M
0 280 V
681 251 M
62 0 V
681 531 M
62 0 V
805 251 M
0 248 V
774 251 M
62 0 V
774 499 M
62 0 V
62 -47 R
0 347 V
867 452 M
62 0 V
867 799 M
62 0 V
990 251 M
0 233 V
959 251 M
62 0 V
959 484 M
62 0 V
62 -233 R
0 218 V
1052 251 M
62 0 V
-62 218 R
62 0 V
62 -174 R
0 221 V
1145 295 M
62 0 V
-62 221 R
62 0 V
62 -139 R
0 189 V
1238 377 M
62 0 V
-62 189 R
62 0 V
62 -201 R
0 188 V
1331 365 M
62 0 V
-62 188 R
62 0 V
61 -205 R
0 158 V
1423 348 M
62 0 V
-62 158 R
62 0 V
62 -132 R
0 158 V
1516 374 M
62 0 V
-62 158 R
62 0 V
62 -244 R
0 172 V
1609 288 M
62 0 V
-62 172 R
62 0 V
62 -185 R
0 166 V
1702 275 M
62 0 V
-62 166 R
62 0 V
62 -133 R
0 125 V
1795 308 M
62 0 V
-62 125 R
62 0 V
61 -58 R
0 124 V
1887 375 M
62 0 V
-62 124 R
62 0 V
62 -171 R
0 111 V
1980 328 M
62 0 V
-62 111 R
62 0 V
62 -136 R
0 111 V
2073 303 M
62 0 V
-62 111 R
62 0 V
62 -48 R
0 111 V
2166 366 M
62 0 V
-62 111 R
62 0 V
62 -213 R
0 125 V
2259 264 M
62 0 V
-62 125 R
62 0 V
LT0
2033 1097 M
180 0 V
499 388 M
37 28 V
37 -6 V
37 -13 V
37 -9 V
37 -4 V
37 1 V
37 0 V
38 -1 V
37 1 V
37 1 V
37 1 V
37 -1 V
37 0 V
37 3 V
37 1 V
37 -2 V
38 2 V
37 -2 V
37 0 V
37 3 V
37 -1 V
37 1 V
37 -1 V
37 0 V
38 1 V
37 -2 V
37 2 V
37 -1 V
37 0 V
37 0 V
37 1 V
37 0 V
38 -2 V
37 2 V
37 0 V
37 -2 V
37 1 V
37 -1 V
37 1 V
37 -2 V
37 1 V
38 1 V
37 0 V
37 -1 V
37 0 V
37 0 V
37 1 V
37 0 V
37 0 V
LT1
2033 997 M
180 0 V
499 385 M
37 33 V
37 -6 V
37 -15 V
37 -11 V
37 -3 V
37 0 V
37 0 V
38 -1 V
37 3 V
37 0 V
37 2 V
37 -1 V
37 -1 V
37 4 V
37 1 V
37 -2 V
38 2 V
37 1 V
37 -3 V
37 3 V
37 0 V
37 0 V
37 -1 V
37 0 V
38 1 V
37 -2 V
37 2 V
37 -1 V
37 0 V
37 0 V
37 1 V
37 -1 V
38 -1 V
37 2 V
37 -1 V
37 -1 V
37 0 V
37 1 V
37 0 V
37 -1 V
37 0 V
38 1 V
37 0 V
37 -1 V
37 1 V
37 0 V
37 -1 V
37 0 V
37 1 V
LT2
2033 897 M
180 0 V
499 393 M
37 8 V
37 -36 V
37 -11 V
37 12 V
37 10 V
37 13 V
37 3 V
38 1 V
37 6 V
37 -3 V
37 -1 V
37 -1 V
37 2 V
37 -2 V
37 3 V
37 -4 V
38 2 V
37 -1 V
37 -2 V
37 1 V
37 0 V
37 0 V
37 -1 V
37 -2 V
38 2 V
37 0 V
37 0 V
37 0 V
37 0 V
37 -1 V
37 1 V
37 -1 V
38 -1 V
37 2 V
37 -2 V
37 0 V
37 1 V
37 -2 V
37 1 V
37 0 V
37 -1 V
38 1 V
37 0 V
37 0 V
37 0 V
37 0 V
37 1 V
37 -1 V
37 -1 V
LT3
2033 797 M
180 0 V
499 394 M
37 8 V
37 -30 V
37 -7 V
37 13 V
37 12 V
37 4 V
37 3 V
38 2 V
37 -5 V
37 1 V
37 -4 V
37 -1 V
37 2 V
37 -2 V
37 4 V
37 -4 V
38 3 V
37 -4 V
37 1 V
37 0 V
37 0 V
37 2 V
37 -2 V
37 0 V
38 2 V
37 -2 V
37 0 V
37 0 V
37 0 V
37 0 V
37 0 V
37 0 V
38 1 V
37 0 V
37 -2 V
37 1 V
37 0 V
37 0 V
37 0 V
37 -1 V
37 1 V
38 0 V
37 0 V
37 -1 V
37 0 V
37 1 V
37 0 V
37 0 V
37 -1 V
LT0
480 528 M
464 0 V
0 -277 V
stroke
grestore
end
showpage
}
\put(1973,797){\makebox(0,0)[r]{$\BE_m$}}
\put(1973,897){\makebox(0,0)[r]{$\BE_\lambda$}}
\put(1973,997){\makebox(0,0)[r]{$\BE_{32}$}}
\put(1973,1097){\makebox(0,0)[r]{$\BE_3$}}
\put(1973,1197){\makebox(0,0)[r]{{\small DELPHI data}}}
\put(1408,1460){\makebox(0,0){(a)}}
\put(1408,51){\makebox(0,0){$x_p$}}
\put(2336,151){\makebox(0,0){0.1}}
\put(1965,151){\makebox(0,0){0.08}}
\put(1594,151){\makebox(0,0){0.06}}
\put(1222,151){\makebox(0,0){0.04}}
\put(851,151){\makebox(0,0){0.02}}
\put(480,151){\makebox(0,0){0}}
\put(420,1221){\makebox(0,0)[r]{600}}
\put(420,944){\makebox(0,0)[r]{400}}
\put(420,667){\makebox(0,0)[r]{200}}
\put(420,390){\makebox(0,0)[r]{0}}
\end{picture}
    \input figure5.tex
    }
    \caption[dummy]{{\it The difference in $x_p$ distributions of
        hadronic W decays between fully hadronic and semi-leptonic
        \ee2ww\ events. Data from \cite{someWWBE}. (b) is a detail
        view of the region close to the origin in (a).}}
    \label{fig:xpww}
  \end{center}
\end{figure}

Comparing fully hadronic and semi-leptonic \ee2ww\ events, one can
also find other observables which may be influenced by BE, and other
interconnection effects between the two W systems. In \cite{someWWBE}
DELPHI found a hint of enhancement in charged multiplicity of fully
hadronic events as compared with twice the multiplicity of isolated W
decays. Also they found an indication of an increase in the
multiplicity for small momentum fractions $x_p=2p_h/E_{\mbox{\tiny
    CM}}$ of the hadrons. Both of these results could be signals for
BE 'cross-talk' between the W's, but at present the errors are much too
large to allow for any conclusions.

In Fig.~\ref{fig:xpww} we present the predictions for the difference in
$x_p$ distributions between \ee2ww\ events with and without cross-talk
for our different algorithms. We see a small effect in the
multiplicity at small $x_p$. However since the local reweighting
scenario conserves the total multiplicity, any enhancement must be
compensated, and this is also predominantly done at small
$x_p$. The difference between the hadronic and leptonic W decays
is therefore the result of a subtraction between almost equally 
large numbers, thereby emphasizing details of the algorithms.

Thus, in all our algorithms, the energy-momentum conservation procedure
reduces the effect of BE enhancement in the $x_p$ spectra at small
$x_p$. Indeed the enhancements are in all cases much smaller than
is indicated by data. Given the large experimental errors we do not take 
this seriously, in particular since L3 and OPAL do not confirm 
the DELPHI observation \cite{noxp}. However, should the signal in 
\cite{someWWBE} survive an increase in statistics, it would not 
necessarily rule out our local reweighting approach as such, but need 
only indicate that we still have a problem with the approach to the 
energy-momentum conservation issue. A difference at small $x_p$ could 
also be caused by other physics mechanisms, such as colour rearrangement 
\cite{TSVK,rearrange}. 

\begin{table}[t]
  \begin{center}
    \begin{tabular}{|l||r|r|r|r|r|}
      \hline model & $\langle\delta m_\W\rangle$ & $\delta\langle
      m_{\W}^{4j\mbox{{\scriptsize 0}}}\rangle$ & $\delta\langle
      m_{\W}^{4j\mbox{{\scriptsize A}}}\rangle$ & $\delta\langle
      m_{\W}^{4j\mbox{{\scriptsize B}}}\rangle$ &
      $\delta\langle m_{\W}^{4j\mbox{{\scriptsize C}}}\rangle$ \\
      & $\pm 1$ & $\pm 4$ & $\pm 8$ & $\pm 8$ & $\pm 8$ \\
      \hline
      {\small (170 GeV)} & & & & & \\
      $\BE_0$ & 130 & & & & \\
      $\BE_3$ & $-8$ & $-6$ & $-4$ & 1 & $-6$ \\
      $\BE_{32}$ & $-9$ & $-8$ & $-3$ & $-5$ & $-2$ \\
      $\BE_\lambda$ & 38 & 38 & 16 & 15 & 12 \\
      $\BE_m$ & 75 & 69 & 15 & 13 & 14 \\
      $\BE_m'$  & 59 & 50 & 2 & 8 & $-5$ \\
      $\BE_m''$ & 102 & 93 & 26 & 25 & 23 \\
      $\BE_m^L$ &  60 & 44 & 17 & 19 & 11 \\
      $\BE_m^{\mbox{\scriptsize peak}}$ & 75 & 70 & 18 & 13 & 16 \\
      \hline
      {\small (190 GeV)} & & & & & \\
      $\BE_m$ & 183 & 191 & 23 & 25 & 14 \\
      $\BE_m'$ & 127 & 114 & $-8$ & $-14$ & $-8$  \\
      \hline
    \end{tabular}
  \end{center}
  \caption[dummy]{{\it Shifts in MeV of the measured mass \tmw4j\
      for different models and different mass reconstruction methods.
      The top number in each column indicates the statistical error
      for a simulated sample of $4\times 10^5$ events. The event
      samples were generated at 170 GeV center of mass energy (except
      the last two rows which were generated at 190 GeV) and the fits
      were restricted to $78.25 < \mw4j < 82.25$ GeV (except the row
      $\BE_m^{\mbox{\scriptsize peak}}$ uses $79.2 < \mw4j < 81.3$
      GeV).}}
  \label{tab:dmw}
\end{table}

We now proceed to estimate the BE-induced shift in the measured W mass
\tmw4j. Since our algorithms preserve the notion that each particle
belongs to a given W, it is easy to obtain a shift in each event by
simply calculating the invariant mass of the decay products of each W
before and after the BE algorithm. The average shift is presented in
table \ref{tab:dmw} in the column denoted $\langle\delta\mw4j\rangle$.
It is clear that the shifts obtained with the new algorithms are
smaller than our previous result in \cite{BE95}\footnote{The
  corresponding value in \cite{BE95} is somewhat lower due to a minor
  error in the averaging procedure.}. This is to be expected as the
energy is conserved locally by pushing pairs of particles away from
each other, counteracting the BE-induced shift. Especially in the
$\BE_3$ and $\BE_{32}$ schemes, the opposing shift is calculated
between the same particles as are affected by the BE shifts, and it is
not surprising that the total shift in the W mass is close to zero.
Put another way, we have previously argued, on physics grounds, that
weights above unity naturally leads to a positive W mass shift, and it
follows in the same spirit that weights below unity gives a negative W
mass shift.  In the $\BE_3$ and $\BE_{32}$ schemes, weights above and
below unity are tuned in such a way that their net effect is expected
to cancel, exactly for energy and approximately for the W mass.

In table~\ref{tab:dmw} we also present the result for some variations
of the $\BE_m$ scheme. $\BE_m'$ is explained below. For $\BE_m''$, if
a pair of identical bosons come from the same W, only pairs of
particles from this W are considered for the energy compensating
shift. In $\BE_m^L$, the shifts $\delta\bfp_i^j$ and $\delta\bfr_i^j$
are calculated in the center of mass system of each pair instead of in
the lab system. In both these cases the changes are moderate and
remind us that there are uncertainties due to the details in the
implementation.

It has been noted that a real measurement of \tmw4j\ would mostly be
sensitive to the peak position of the mass distribution, and in
\cite{Rasmus} it was found that the small BE-induced shift in
$\langle\delta\mw4j\rangle$ mostly stem from the tails of the
distribution. The BE shift thus almost disappears if the mass is
obtained from a fit to a relativistic Breit--Wigner (plus background).
Doing the same with our algorithms we find no significant decrease of
the BE shift, however, as seen in the column denoted $\delta\langle
m_{\W}^{4j\mbox{{\scriptsize 0}}}\rangle$ in table \ref{tab:dmw}. It
is possible this partly comes from the difference between models with
global weights and those without. Specifically, if the average value
of the global weight has a nontrivial energy dependence, then the
weighting procedure would skew the wings.  However, this is just a
guess, and further studies are required to settle the issue.

It is clear that the mass shift in our algorithms would mostly come
from the softest particles in the events. These are also the ones that
are most difficult to associate to one or the other of $\W^+$ and
$\W^-$. To achieve a more experimental-like situation we therefore
ignore what the generator tells us about the origin of each
final-state particle and instead perform a jet clustering in the same
way as in \cite{TSVK}.  Three different strategies are studied for
associating jets with either W boson, denoted A, B, and C in
table~\ref{tab:dmw}. In all cases the {\sc LUCLUS} jet clustering
algorithm \cite{pythia} is used to reconstruct exactly four jets.
These are then paired together to represent a $\W^+$ and a $\W^-$. In
each event the combination $(j_1j_2)(j_3j_4)$ is chosen which
minimises $|m_{j_1j_2}-80|+|m_{j_3j_4}-80|$ (A) or
$|m_{j_1j_2}-80+m_{j_3j_4}-80|$ (C) or maximizes the angles between
the jets $\theta_{j_1j_2}+\theta_{j_3j_4}$ (B). The reconstructed mass
distribution is thereafter again fitted to extract a peak position.

\begin{figure}
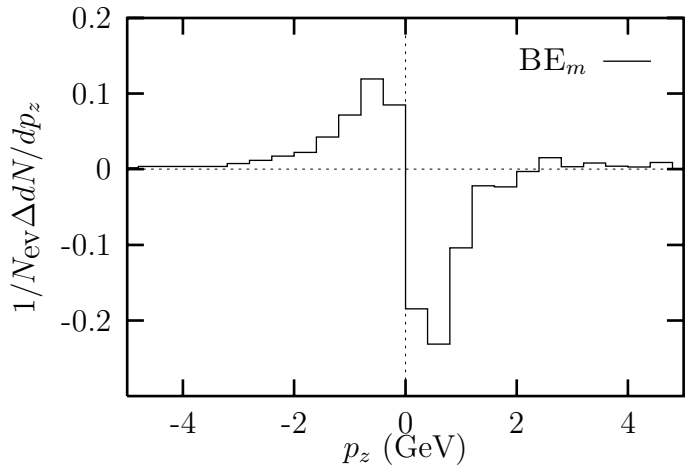

  \begin{center}
    \input figure6.tex
    \caption[dummy]{{\it The difference between the $p_z$ distribution
        with and without BE correlations between the $\W$'s according
        to the $\BE_m$ algorithm, for events with
        $|p_{zq1}-p_{zq2}|+|p_{zq3}-p_{zq4}|<E_{\mbox{cm}}/2$. $p_z$
        for a particle is the momentum component along the direction
        of the $\W$ from which it was produced.}}
    \label{fig:dpz}
  \end{center}
\end{figure}

In all cases the BE-induced shift is reduced. It seems that the
BE-shifts increases the likelihood that soft particles become
misassigned in such a way that the momenta of the W's are increased.
(We remind that, by energy conservation, an increased W momentum
corresponds to a decreased W mass.) To see how this can come about,
assume that the four jets of an event separate into one $\W^+$ and one
$\W^-$ hemisphere, i.e that the two jets of the $\W^+$ ($\W^-$) have a
positive longitudinal momentum with respect to the $\W^+$ ($\W^-$)
direction of motion. Stray particles in the `wrong' hemisphere would
then have a large likelihood of being misassigned. Such a
misassignment removes particles with momentum opposite to the motion
of the W itself and adds them to the other W, thus increasing the
reconstructed momentum of both. Since our implementation of BE effects
tends to enhance particle production in the central region of the
event and particularly the migration of particles in the direction of
the other W, we would then expect an effect of the observed sign. When
the jets of a W are not in the same hemisphere, the effects of
misassignments could more easily go either way, so the influence on
the W mass should be reduced. To quantify effects, consider events 
aligned with the $\W^+$ along the $+z$ axis and then require
$\delta p_z = |p_{zq_1}-p_{z\bar{q}_2}| + |p_{zq_3}-p_{z\bar{q}_4}| <
E_{\mbox{\scriptsize CM}}/2$, using generator information about the
$z$-components of the initial quarks from the W decays
($\W^+\rightarrow q_1\bar{q}_2$, $\W^-\rightarrow q_3\bar{q}_4$).
Using a simple cut at $p_z=0$ we can get an estimate of the BE-induced
misassignment effects by studying the difference in $p_z$ distribution
of particles from one W with and without BE-cross-talk.  The result is
shown in Fig.~\ref{fig:dpz} for the $\BE_m$ algorithm and we see that
the misassignment is indeed increased. Integrating the curve in
Fig.~\ref{fig:dpz} we find an average increase in the W momentum of
the order of 100 MeV, which would correspond to a shift in the
reconstructed W mass of about $-40$~MeV. Note that this shift is
negative, so the statement in our previous publication \cite{BE95}
that BE effects necessarily would increase the measured mass in not
quite true. Note also that one could imagine that BE effects in this
way could affect the measured mass even if the actual W masses are
unaffected. For the $\BE_{32}$ algorithm, however, the BE-induced
misassignment effects are much smaller and we see no effect for the A,
B and C reconstruction in table~\ref{tab:dmw}.

\begin{table}[t]
  \begin{center}
    \begin{tabular}{|l||r|r|r|}
      \hline
      shift & low & medium & high \\
      \hline
      $\Delta\langle m_{\W}^{4j\mbox{{\scriptsize A}}}\rangle$ no $\BE$ &
      -62 & +175 & +189 \\
      $\delta\langle m_{\W}^{4j\mbox{{\scriptsize 0}}}\rangle$ $\BE_m$ &
      +88 & +64 & +52 \\
      $\Delta\langle m_{\W}^{4j\mbox{{\scriptsize A}}}\rangle$ $\BE_m$ &
      -137 & +139 & +134 \\
      $\delta\langle m_{\W}^{4j\mbox{{\scriptsize A}}}\rangle$ $\BE_m$ &
      +13 & +28 & -3 \\
      \hline      
    \end{tabular}
  \end{center}
  \caption[dummy]{{\it Shifts in the W mass peak position due to
      reconstruction and BE effects for different topologies. Low
      means $\delta p_z<E_{\mbox{\scriptsize CM}}*0.4$, high means
      $\delta p_z>E_{\mbox{\scriptsize CM}}*0.6$. $\Delta\langle
      m_{\W}^{4j\mbox{{\scriptsize A}}}\rangle$ is the shift in the
      peak position due to the reconstruction, while $\delta\langle
      m_{\W}^{4j\mbox{{\scriptsize 0}}}\rangle$ and $\delta\langle
      m_{\W}^{4j\mbox{{\scriptsize A}}}\rangle$ are defined as for
      table \ref{tab:dmw}. The statistical error is everywhere around
      10 MeV. Note the relation 
      $\delta\langle m_{\W}^{4j\mbox{{\scriptsize A}}}\rangle =
       \delta\langle m_{\W}^{4j\mbox{{\scriptsize 0}}}\rangle +
       \Delta\langle m_{\W}^{4j\mbox{{\scriptsize A}}}\rangle
       (\mbox{BE} - \mbox{no~BE})$,
      i.e. the observable W mass shift by BE effects is the sum of the
      theoretical mass shift for `correct' assignment of particles and
      the mass shift by `erroneous' particle assignments when moving from
      the no-BE to the BE world.}}
  \label{tab:dpz}
\end{table}

Looking more closely at the effects of reconstruction method A, we see
in table \ref{tab:dpz} that without any BE cross-talk, the measured W
mass is affected differently for different event topologies (again
using $\delta p_z$ above as a topology measure). For small $\delta
p_z$ the mass is shifted downwards, while for larger $\delta p_z$, the
shift is positive. In table \ref{tab:dpz} we also see that the {\em
  direct} BE shift is positive everywhere, although largest at small
$\delta p_z$.  But with BE cross-talk, the reconstruction effects are
also changed, and the reconstructed mass is lowered everywhere as
compared with the case of no cross-talk. At small $\delta p_z$, where
the direct BE shift is largest, the additional negative shift due to
BE-induced reconstruction effects is also larger, and everywhere the
direct BE shift is more or less compensated by BE effects in the
reconstruction.

\begin{figure}
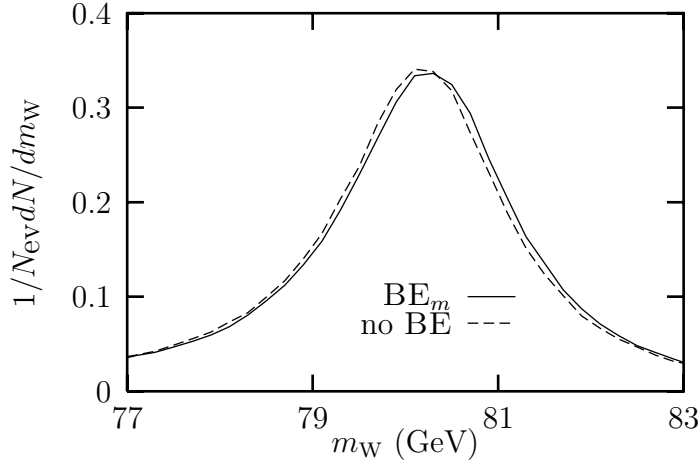

  \begin{center}
    \input figure7.tex
    \caption[dummy]{{\it Shape of the W mass peak with and
       without BE included according to model $\BE_m$.}}
    \label{fig:masspeak}
  \end{center}
\end{figure}

Above we noted the increase in fitted W width in some global weight 
models. Also in our models is the width increased by BE effects, 
table~\ref{tab:wid}. The order of the width increase is 40~MeV, i.e. 
comparable with what is found in the global models. In retrospect, a 
broadening of the W peak is a not unnatural consequence of the 
fluctuations in the BE-induced W mass shifts. 
That is, till now we have discussed the shift of the {\em average} W 
mass in a large event sample. The shift in each {\em individual} W mass 
is much larger, typically 200~MeV, cf. table \ref{tab:wid}. This
variability is rather weakly correlated with the W mass itself, but
is instead mainly given by the W decay angles and
fluctuations in the fragmentation process. The observable W width 
is therefore increased in relation to the width of the BE mass shift 
distribution. A crude addition in quadrature gives the right order 
of magnitude of the effects,
$\delta \langle \Gamma_{\W} \rangle \sim 
\Gamma_{\mbox{\scriptsize BE}}^2 / 2 \langle \Gamma_{\W} \rangle \sim
2 \sigma_{\mbox{\scriptsize BE}}^2 / \langle \Gamma_{\W} \rangle$.
One should note, however, that the error on the W width determination
is rather large, so it is doubtful whether a 40~MeV increase in the
W width will be observable at LEP~2. Specifically, our models give 
only a very modest drop of peak height, Fig.~\ref{fig:masspeak}, and 
the total cross section in the central peak is essentially unchanged.
This should be contrasted with the model of Fia{\l}kowski and Wit,
where there is a significant increase of the low-mass tail, beyond 
the range of the W peak fit, and a corresponding drop of the peak
value. Whereas thus an increase of the W width seems to be a common
phenomenon in many models, the difference is whether this is mainly
a broadening of the central mass peak or also has significant 
implications for the wings.

\begin{table}[t]
  \begin{center}
    \begin{tabular}{|l||r|r|r|r|r|}
      \hline model & $\sigma_{\mbox{\scriptsize BE}}$ & $\delta\langle
      \Gamma_{\W}^{4j\mbox{{\scriptsize 0}}}\rangle$ & $\delta\langle
      \Gamma_{\W}^{4j\mbox{{\scriptsize A}}}\rangle$ & $\delta\langle
      \Gamma_{\W}^{4j\mbox{{\scriptsize B}}}\rangle$ &
      $\delta\langle \Gamma_{\W}^{4j\mbox{{\scriptsize C}}}\rangle$ \\
      & & $\pm 10$ & $\pm 31$ & $\pm 34$ & $\pm 28$ \\
      \hline
      $\BE_3$ & 36 & 6 & 44 & 49 & 49 \\
      $\BE_{32}$ & 47 & 8 & 28 & 27 & 39 \\
      $\BE_\lambda$ & 250 & 80 & 48 & 36 & 29 \\
      $\BE_m$ & 190 & 34 & 44 & 39 & 42 \\
      $\BE_m'$  & 180 & 31 & 66 & 76 & 70 \\
      $\BE_m''$ & 140 & 6 & 54 & 51 & 44 \\
      $\BE_m^L$ & 170 & 29 & 28 & 24 & 30 \\
      $\BE_m^{\mbox{\scriptsize peak}}$ & 190 & 56 & 48 & 49 & 28 \\
      \hline
    \end{tabular}
  \end{center}
  \caption[dummy]{{\it The fitted width for different models and different
      mass reconstruction methods. Notation as in table \ref{tab:dmw}.
      Also shown is $\sigma_{\mbox{\scriptsize BE}}$, the Gaussian width of
      the true BE-induced mass shift.}}
  \label{tab:wid}
\end{table}

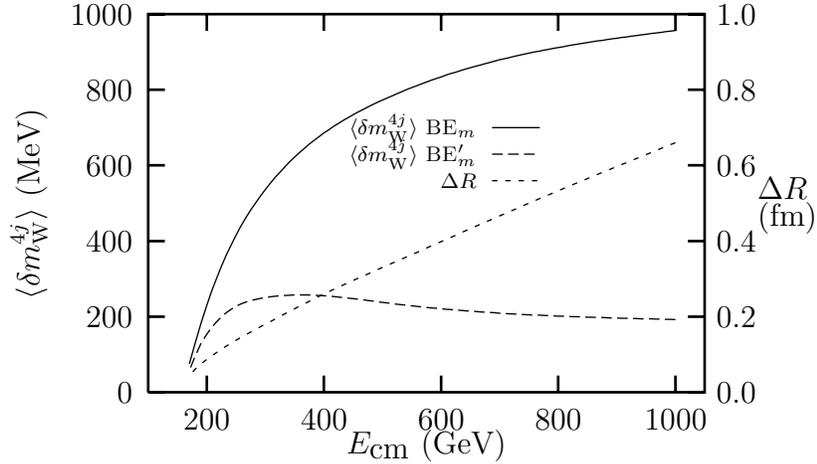
\begin{figure}
  \begin{center}
\setlength{\unitlength}{0.1bp}
\special{!
/gnudict 40 dict def
gnudict begin
/Color false def
/Solid false def
/gnulinewidth 5.000 def
/vshift -33 def
/dl {10 mul} def
/hpt 31.5 def
/vpt 31.5 def
/M {moveto} bind def
/L {lineto} bind def
/R {rmoveto} bind def
/V {rlineto} bind def
/vpt2 vpt 2 mul def
/hpt2 hpt 2 mul def
/Lshow { currentpoint stroke M
  0 vshift R show } def
/Rshow { currentpoint stroke M
  dup stringwidth pop neg vshift R show } def
/Cshow { currentpoint stroke M
  dup stringwidth pop -2 div vshift R show } def
/DL { Color {setrgbcolor Solid {pop []} if 0 setdash }
 {pop pop pop Solid {pop []} if 0 setdash} ifelse } def
/BL { stroke gnulinewidth 2 mul setlinewidth } def
/AL { stroke gnulinewidth 2 div setlinewidth } def
/PL { stroke gnulinewidth setlinewidth } def
/LTb { BL [] 0 0 0 DL } def
/LTa { AL [1 dl 2 dl] 0 setdash 0 0 0 setrgbcolor } def
/LT0 { PL [] 0 1 0 DL } def
/LT1 { PL [4 dl 2 dl] 0 0 1 DL } def
/LT2 { PL [2 dl 3 dl] 1 0 0 DL } def
/LT3 { PL [1 dl 1.5 dl] 1 0 1 DL } def
/LT4 { PL [5 dl 2 dl 1 dl 2 dl] 0 1 1 DL } def
/LT5 { PL [4 dl 3 dl 1 dl 3 dl] 1 1 0 DL } def
/LT6 { PL [2 dl 2 dl 2 dl 4 dl] 0 0 0 DL } def
/LT7 { PL [2 dl 2 dl 2 dl 2 dl 2 dl 4 dl] 1 0.3 0 DL } def
/LT8 { PL [2 dl 2 dl 2 dl 2 dl 2 dl 2 dl 2 dl 4 dl] 0.5 0.5 0.5 DL } def
/P { stroke [] 0 setdash
  currentlinewidth 2 div sub M
  0 currentlinewidth V stroke } def
/D { stroke [] 0 setdash 2 copy vpt add M
  hpt neg vpt neg V hpt vpt neg V
  hpt vpt V hpt neg vpt V closepath stroke
  P } def
/A { stroke [] 0 setdash vpt sub M 0 vpt2 V
  currentpoint stroke M
  hpt neg vpt neg R hpt2 0 V stroke
  } def
/B { stroke [] 0 setdash 2 copy exch hpt sub exch vpt add M
  0 vpt2 neg V hpt2 0 V 0 vpt2 V
  hpt2 neg 0 V closepath stroke
  P } def
/C { stroke [] 0 setdash exch hpt sub exch vpt add M
  hpt2 vpt2 neg V currentpoint stroke M
  hpt2 neg 0 R hpt2 vpt2 V stroke } def
/T { stroke [] 0 setdash 2 copy vpt 1.12 mul add M
  hpt neg vpt -1.62 mul V
  hpt 2 mul 0 V
  hpt neg vpt 1.62 mul V closepath stroke
  P  } def
/S { 2 copy A C} def
end
}
\begin{picture}(2880,1728)(0,0)
\special{"
gnudict begin
gsave
50 50 translate
0.100 0.100 scale
0 setgray
/Helvetica findfont 100 scalefont setfont
newpath
-500.000000 -500.000000 translate
LTa
600 251 M
2097 0 V
LTb
600 251 M
63 0 V
2034 0 R
-63 0 V
600 536 M
63 0 V
2034 0 R
-63 0 V
600 821 M
63 0 V
2034 0 R
-63 0 V
600 1107 M
63 0 V
2034 0 R
-63 0 V
600 1392 M
63 0 V
2034 0 R
-63 0 V
600 1677 M
63 0 V
2034 0 R
-63 0 V
821 251 M
0 63 V
0 1363 R
0 -63 V
1262 251 M
0 63 V
0 1363 R
0 -63 V
1704 251 M
0 63 V
0 1363 R
0 -63 V
2145 251 M
0 63 V
0 1363 R
0 -63 V
2587 251 M
0 63 V
0 1363 R
0 -63 V
600 251 M
2097 0 V
0 1426 V
-2097 0 V
600 251 L
LT0
1896 1249 M
180 0 V
511 366 R
-19 -2 V
-18 -2 V
-19 -3 V
-18 -2 V
-18 -2 V
-19 -3 V
-18 -2 V
-19 -3 V
-18 -2 V
-18 -3 V
-19 -2 V
-18 -3 V
-18 -2 V
-19 -3 V
-18 -3 V
-19 -2 V
-18 -3 V
-18 -3 V
-19 -3 V
-18 -3 V
-19 -3 V
-18 -3 V
-18 -4 V
-19 -3 V
-18 -3 V
-19 -4 V
-18 -3 V
-18 -4 V
-19 -3 V
-18 -4 V
-19 -4 V
-18 -4 V
-18 -4 V
-19 -5 V
-18 -4 V
-19 -4 V
-18 -5 V
-18 -5 V
-19 -5 V
-18 -5 V
-19 -5 V
-18 -5 V
-18 -5 V
-19 -6 V
-18 -6 V
-19 -5 V
-18 -6 V
-18 -7 V
-19 -6 V
-18 -6 V
-18 -7 V
-19 -7 V
-18 -7 V
-19 -7 V
-18 -7 V
-18 -8 V
-19 -8 V
-18 -8 V
-19 -8 V
-18 -8 V
-18 -8 V
-19 -9 V
-18 -9 V
-19 -10 V
-18 -9 V
-18 -10 V
-19 -11 V
-18 -10 V
-19 -12 V
-18 -11 V
-18 -13 V
-19 -12 V
-18 -14 V
-19 -14 V
-18 -14 V
-18 -16 V
-19 -16 V
-18 -17 V
-19 -18 V
-18 -19 V
-18 -20 V
-19 -20 V
-18 -22 V
-19 -24 V
-18 -24 V
-18 -26 V
986 938 L
968 909 L
949 877 L
931 843 L
913 807 L
894 768 L
876 726 L
858 681 L
839 632 L
821 579 L
799 511 L
777 436 L
755 358 L
LT1
1896 1149 M
180 0 V
2587 525 M
-19 1 V
-18 0 V
-19 1 V
-18 0 V
-18 1 V
-19 0 V
-18 0 V
-19 1 V
-18 0 V
-18 1 V
-19 0 V
-18 1 V
-18 0 V
-19 1 V
-18 1 V
-19 0 V
-18 1 V
-18 0 V
-19 1 V
-18 1 V
-19 0 V
-18 1 V
-18 1 V
-19 0 V
-18 1 V
-19 1 V
-18 1 V
-18 1 V
-19 0 V
-18 1 V
-19 1 V
-18 1 V
-18 1 V
-19 1 V
-18 1 V
-19 1 V
-18 2 V
-18 1 V
-19 1 V
-18 1 V
-19 2 V
-18 1 V
-18 1 V
-19 2 V
-18 1 V
-19 2 V
-18 1 V
-18 2 V
-19 2 V
-18 1 V
-18 2 V
-19 2 V
-18 2 V
-19 2 V
-18 2 V
-18 2 V
-19 2 V
-18 2 V
-19 3 V
-18 2 V
-18 2 V
-19 3 V
-18 2 V
-19 3 V
-18 2 V
-18 2 V
-19 2 V
-18 2 V
-19 2 V
-18 2 V
-18 2 V
-19 1 V
-18 1 V
-19 1 V
-18 1 V
-18 0 V
-19 0 V
-18 0 V
-19 -1 V
-18 -1 V
-18 -1 V
-19 -2 V
-18 -2 V
-19 -3 V
-18 -3 V
-18 -4 V
-19 -4 V
-18 -6 V
-19 -7 V
-18 -9 V
913 564 L
894 551 L
876 535 L
858 517 L
839 495 L
821 470 L
799 435 L
777 387 L
755 333 L
LT2
1896 1049 M
180 0 V
511 143 R
-19 -7 V
-18 -8 V
-19 -7 V
-18 -8 V
-18 -7 V
-19 -8 V
-18 -7 V
-19 -8 V
-18 -7 V
-18 -8 V
-19 -7 V
-18 -8 V
-18 -7 V
-19 -8 V
-18 -7 V
-19 -8 V
-18 -8 V
-18 -7 V
-19 -8 V
-18 -7 V
-19 -8 V
-18 -8 V
-18 -7 V
-19 -8 V
-18 -8 V
-19 -8 V
-18 -7 V
-18 -8 V
-19 -8 V
-18 -8 V
-19 -8 V
-18 -8 V
-18 -8 V
-19 -8 V
-18 -8 V
-19 -8 V
-18 -8 V
-18 -8 V
-19 -8 V
-18 -8 V
-19 -8 V
-18 -8 V
-18 -8 V
-19 -8 V
-18 -9 V
-19 -8 V
-18 -8 V
-18 -8 V
-19 -8 V
-18 -8 V
-18 -8 V
-19 -9 V
-18 -8 V
-19 -8 V
-18 -8 V
-18 -8 V
-19 -8 V
-18 -8 V
-19 -8 V
-18 -8 V
-18 -8 V
-19 -9 V
-18 -8 V
-19 -8 V
-18 -8 V
-18 -8 V
-19 -8 V
-18 -9 V
-19 -8 V
-18 -9 V
-18 -8 V
-19 -9 V
-18 -9 V
-19 -9 V
-18 -10 V
-18 -9 V
-19 -9 V
-18 -10 V
-19 -10 V
-18 -9 V
-18 -10 V
-19 -10 V
-18 -10 V
-19 -9 V
-18 -10 V
-18 -10 V
986 478 L
968 468 L
949 458 L
931 447 L
913 436 L
894 425 L
876 413 L
858 401 L
839 388 L
821 375 L
799 358 L
777 338 L
755 314 L
stroke
grestore
end
showpage
}
\put(1836,1049){\makebox(0,0)[r]{{\scriptsize $\Delta R$}}}
\put(1836,1149){\makebox(0,0)[r]{{\scriptsize $\langle\delta\mw4j\rangle$ $\BE_m'$}}}
\put(1836,1249){\makebox(0,0)[r]{{\scriptsize $\langle\delta\mw4j\rangle$ $\BE_m$}}}
\put(2896,907){\makebox(0,0)[l]{(fm)}}
\put(2896,1021){\makebox(0,0)[l]{$\Delta R$}}
\put(2741,251){\makebox(0,0)[l]{0.0}}
\put(2741,536){\makebox(0,0)[l]{0.2}}
\put(2741,821){\makebox(0,0)[l]{0.4}}
\put(2741,1107){\makebox(0,0)[l]{0.6}}
\put(2741,1392){\makebox(0,0)[l]{0.8}}
\put(2741,1677){\makebox(0,0)[l]{1.0}}
\put(1648,51){\makebox(0,0){$E_{\mbox{cm}}$ (GeV)}}
\put(220,964){%
\special{ps: gsave currentpoint currentpoint translate
270 rotate neg exch neg exch translate}%
\makebox(0,0)[b]{\shortstack{$\langle\delta\mw4j\rangle$ (MeV)}}%
\special{ps: currentpoint grestore moveto}%
}
\put(2587,151){\makebox(0,0){1000}}
\put(2145,151){\makebox(0,0){800}}
\put(1704,151){\makebox(0,0){600}}
\put(1262,151){\makebox(0,0){400}}
\put(821,151){\makebox(0,0){200}}
\put(540,1677){\makebox(0,0)[r]{1000}}
\put(540,1392){\makebox(0,0)[r]{800}}
\put(540,1107){\makebox(0,0)[r]{600}}
\put(540,821){\makebox(0,0)[r]{400}}
\put(540,536){\makebox(0,0)[r]{200}}
\put(540,251){\makebox(0,0)[r]{0}}
\end{picture}
    \caption[dummy]{{\it The average shift in MeV of the measured mass
        \tmw4j\ for the $\BE_m$ and $\BE_m'$ models as a
        function of the $\e^+\e^-$ center of mass energy (in GeV).
        Also shown is the average separation $\Delta R$ in fm between the
        decay vertices of $\W^+$ and $\W^-$.}}
    \label{fig:dmwe}
  \end{center}
\end{figure}

In \cite{BE95} we noted that the shift in \tmw4j\ increases with the
center of mass energy, and explained why this is a natural behaviour.
This is still true e.g.\ for the $\BE_m$ model, as seen in
Fig.~\ref{fig:dmwe}.  However, the argumentation is based on the
assumption that the fragmentation regions of the $\W^+$ and the $\W^-$
do overlap significantly, as is the case over the LEP~2 energy range.
At very high energies the shift should go away, since here the W's
decay only after they have travelled well apart.  The separation of
the decay vertices can be taken into account, approximately, by using
a modified radius in the \tf2q\ function in equation \ref{qshift} when
calculating shifts for particles from different W bosons (but not from
the same W). Specifically, the procedure described in \cite{TSVK} is
used to generate the distance $\Delta R$ between the decay vertices of
the two W's, based on a Monte Carlo sampling of the expected W decay
distribution as a function of the W mass. We then define a modified
version of $\BE_m$, denoted $\BE_m'$, with
\begin{equation}
  f_2^{+-}(Q)=1+\lambda\exp(Q^2(R+\Delta R)^2)
\end{equation}
for pairs from different W's. In Fig.~\ref{fig:dmwe} it is seen that 
the separation does indeed lower the shift, but the shift still rises 
with the center of mass energy until around 400 GeV,
whereafter it slowly decreases. The effect of the experimental
reconstruction procedures can be seen from table~\ref{tab:dmw},
where the $\BE_m$ model shows the expected increase, while $\BE'_m$ 
remains close to zero and possibly is decreasing towards more negative 
values. The net uncertainty therefore indeed does seem to increase over 
the LEP~2 energy range.

\section{Conclusions}
\label{sec:sum}

This paper has two objectives: to take a critical look at the 
modelling of BE effects, especially for its impact on the W mass,
and to develop improved versions of local weight algorithms. 

Today the `global weight' approach to the BE phenomenon dominates. 
However, many global weight schemes have basic weaknesses, in areas
such as the theoretical one of factorization or the experimental 
ones of comparisons with $\Z^0$ total and partial widths, cross 
sections, jet rates, and so on. Furthermore, one can easily see ways 
to construct global weight models that could give misleading results, 
e.g.\ if the average BE weight has a nontrivial dependence on the 
mass of each W or on the jet topology of the W decays. 
In general, the arbitrariness of the weight rescaling schemes probably 
is the limiting factor when trying to extract reliable predictions out 
of several current global weight algorithms. 
Even when factorization is respected, there is no unique recipe for 
how BE effects could couple the two $\W$ hadronization processes.

Therefore we do not consider the matter settled. The local weight 
approach is certainly not free of objections, but it does address 
and solve some of the basic issues that the pure global weight 
approach does not. However, just as there exist
a multitude of mutually contradictory global models in the literature, 
one can construct many kinds of local models. In this paper we have
come up with four main alternatives to the scheme in \cite{BE95}. 
For technical simplicity, all four are based on the same kind of 
momentum shifting strategy as in the original one, but they are 
still sufficiently different to probe a wide space of local weight 
models. The models are in this paper applied to the topical issue
of the W mass, but clearly can be used also for Z$^0$ physics and
other studies. Therefore, should the experimental verdict be that
no BE effects connect the two W's, the algorithms we have proposed 
here could still be used to explore other aspects of the BE 
phenomenon. Should an effect be found, on the other hand, it would be 
even more interesting to understand whether the algorithms can be 
discriminated by more detailed comparisons with LEP~1 data.

In our original paper \cite{BE95} we stated that the model studied 
there was likely to give an estimate of the maximal possible effects,
with the real ones some unknown fraction thereof. Indeed, the models
studied here at most reach three quarters of the original W mass 
shift, and range down to essentially zero mass shift. This is based
on untuned models, however, and we expect that a careful tuning to 
$\Z^0$ data would bring up the numbers somewhat. Global weight
models cluster around zero. There are exceptions that show
some shift, but none anywhere near as big as our original scenario.
Furthermore, with the new models we now have the possibility to
study the impact of the experimental procedures used to extract a W 
mass. Unlike the results of ref. \cite{Rasmus}, a fit to the peak
position of the W Breit-Wigner does not significantly reduce the 
theoretical mass shift in our models. Instead a reduction occurs by 
another mechanism: the shift of the momenta of particles belonging
to one W in the direction of the other W. So long as these particles
are bookkept with their original W, it is precisely this mechanism
that reduces the W momenta and hence increases the W masses in the 
first place. When particles are shifted so far that they tend to be 
assigned to the `wrong' W, however, the reconstructed momentum of 
each W can instead increase and the W mass shift is thereby reduced.

In the end, we therefore remain with W mass shifts up to at most
30 MeV at 170 GeV. These models still have to be retuned somewhat, 
cf. Fig.~\ref{fig:c2}, and the uncertainty would increase with energy,
but something like 50~MeV seems to be a safe upper limit over the 
LEP~2 energy range. All the numbers here refer to our attempts at
reproducing a sensible experimental procedure. As we have seen,
however, the BE phenomenon does involve low-momentum particle and
contains nontrivial dependences on the event topology, so the only
realistic numbers are those that are obtained by the experimental
collaborations, with their selection cuts and within their acceptance.
Disregarding such issues,
it would be tempting to take some kind of average of the different
model studies, ours and those of others, and claim that the 
uncertainty on the W mass from BE effects is even smaller, maybe
not more than 10--15 MeV. However, nature is not a
democratic compromize between ten models. There exists {\em one}
correct description of BE effects and, if we are honest, we have to
admit that all the models we use are likely to be flawed with respect
to this truth. Therefore an estimate of the uncertainty had better be
based on the most `pessimistic' scenario that is not in blatant
disagreement with existing data. 
 
This does not mean prospects are hopeless. The DELPHI \cite{NoWWBE}
and ALEPH \cite{ALEPHno} studies point the way to constraining the
amount of cross-talk occuring between the $\W^+$ and $\W^-$ hadronic
systems, once the statistics is improved. An observation of no
cross-talk would certainly settle the issue, in the sense that we (at
least currently) do not know of any way to construct a BE model that
would give a $C^*_2(Q) \equiv 1$ (eq.~(\ref{Rstarstar})) and still
induce a W mass shift.  However, note that the converse does not hold:
models with similar nonunity $C^*_2(Q)$ shapes may disagree on the W
mass shift value.  What can be said, however, is that the closer
$C^*_2(Q)$ is constrained to unity, the smaller the maximum imaginable
W mass shift.

While clearly the observation of BE effects spanning the two W's
would be very exciting, also a null result would be very interesting
and in need of an explanation. (How do two hadronizing systems,
that clearly overlap in space and time, manage not to feel each
other?) Continued BE studies therefore are well worth the effort.

\section*{Acknowledgements}

We would like to thank B.~Andersson, K.~Fia{\l}kowski, U.~Heinz,
S.~Jadach, R.~M{\o}ller, M.~Ring\'er, \v{S}.~Todorova--Nov\'a and
A.~Tomaradze for useful discussions. The opinions expressed in this 
paper are our own, however.

\end{document}